\documentclass[a4paper,10pt]{article}
\usepackage[centertags]{amsmath}
\usepackage{amsmath}
\usepackage[dutch,british]{babel}
\usepackage{amsfonts}
\usepackage{amssymb}
\usepackage{amsthm}
\usepackage{newlfont}
\usepackage{color}

\theoremstyle{plain}
  \newtheorem{theorem}{Theorem}[section]
  
  \newtheorem{proposition}[theorem]{Proposition}
  \newtheorem{lemma}[theorem]{Lemma}
  \newtheorem{remark}[theorem]{Remark}
\theoremstyle{definition}
  \newtheorem{definition}{Definition}[section]
  \newtheorem{assumption}[theorem]{Assumption}

\theoremstyle{remark}

\numberwithin{equation}{section}

\newtheorem{vetremark}[theorem]{\bf{Remark}}

\newcommand\otimesal{\mathop{\hbox{\raise 1.6 ex
  \hbox{$\scriptscriptstyle\mathrm{al}$}
\kern -0.92 em \hbox{$\otimes$}}}}
\newcommand\oplusal{\mathop{\hbox{\raise 1.6 ex
  \hbox{$\scriptscriptstyle\mathrm{al}$}
\kern -0.92 em \hbox{$\oplus$}}}}
\newcommand\Gammal{\hbox{\raise 1.7 ex
\hbox{$\scriptscriptstyle\mathrm{al}$}\kern -0.50 em $\Gamma$}}
\renewcommand\i{\mathrm{i}}

 \let\la=\lambda \let\om=\omega

 \let\Ga=\Gamma  \let\Om=\Omega

\newcommand{\caB}{{\mathcal B}}
\newcommand{\caC}{{\mathcal C}}
\newcommand{\caD}{{\mathcal D}}

\newcommand{\caF}{{\mathcal F}}
\newcommand{\caG}{{\mathcal G}}
\newcommand{\caH}{{\mathcal H}}

\newcommand{\caJ}{{\mathcal J}}
\newcommand{\caK}{{\mathcal K}}

\newcommand{\caM}{{\mathcal M}}

\newcommand{\caP}{{\mathcal P}}

\newcommand{\caR}{{\mathcal R}}

\newcommand{\caZ}{{\mathcal Z}}

\newcommand{\bbN}{{\mathbb N}}

\newcommand{\bbR}{{\mathbb R}}

\newcommand{\opunit}{\text{1}\kern-0.22em\text{l}}

\newcommand{\frh}{{\mathfrak h}}

\newcommand{\slim}{{\mathrm s-}\lim}
\newcommand{\e}{{\mathrm e}}
\newcommand{\iu}{{\mathrm i}}
\renewcommand{\d}{{\mathrm d}}

\newcommand{\res}{{\mathrm R}}

\renewcommand{\sp}{\mathrm{sp}}
\newcommand{\Ran}{\mathrm{Ran}}
\newcommand{\Sym}{\mathrm{Sym}}
\newcommand{\Dom}{\mathrm{Dom}}
\newcommand{\Span}{\mathrm{Span}}
\newcommand{\beq}{ \begin{equation} }
\newcommand{\eeq}{ \end{equation} }
\newcommand{\bet}{ \begin{theorem} }
\newcommand{\eet}{ \end{theorem} }

\newcommand{\s}{{\mathrm s}}

\newcommand{\ren}{\mathrm{ren}}

 \newcounter{smallarabics}
\newenvironment{arabicenumerate}
{\begin{list}{{\normalfont\textrm{\arabic{smallarabics})}}}
  {\usecounter{smallarabics}\setlength{\itemindent}{0cm}
  \setlength{\leftmargin}{5ex}\setlength{\labelwidth}{4ex}
  \setlength{\topsep}{0.75\parsep}\setlength{\partopsep}{0ex}
   \setlength{\itemsep}{0ex}}}
{\end{list}}

\newcounter{smallroman}

\newcommand{\ben}{\begin{arabicenumerate}}
\newcommand{\een}{\end{arabicenumerate}}

\newcommand{\Pair}{\mathrm{Pair}}

\newcommand{\sfock}{\Ga_{\mathrm{s}}}

\newcommand{\wdr}{ }

\begin{document}
\begin{center}
\noindent{\large \bf Extended Weak Coupling
 Limit\\ for Pauli-Fierz Operators} \\

\vspace{15pt}

{\bf Jan Derezi\'{n}ski}
\\
 Department of Mathematical Methods in Physics \\
Warsaw University \\
 Ho\.{z}a 74, 00-682, Warszawa, Poland\\
email: {\tt jan.derezinski@fuw.edu.pl}\\
\vspace{10pt}
{\bf Wojciech De
Roeck} \\
Instituut voor Theoretische
Fysica, K.U.Leuven\\
 Belgium \\
 email:
{\tt wojciech.deroeck@fys.kuleuven.be}\\

\end{center}

\vspace{20pt} \footnotesize \noindent {\bf Abstract: } We consider
the weak coupling limit for a quantum system consisting of a small
subsystem and  reservoirs. It is known rigorously since
\cite{davies1} that the Heisenberg evolution restricted to the
small system converges in an appropriate sense to a Markovian
semigroup. In the nineties, Accardi, Frigerio and Lu
\cite{accardifrigerio} initiated an investigation of the
convergence of the unreduced unitary evolution to a singular
unitary evolution generated by a Langevin-Schr\"odinger equation.
 We present a version of this convergence which
is both simpler and stronger than the formulations which we know.
Our main result  says that  in an appropriately understood weak coupling limit
the interaction of the small system with environment can be expressed in terms
of the so-called quantum white noise.

\vspace{5pt}

  \vspace{20pt}
\normalsize

\section{Introduction}

One of the main goals of mathematical physics is to justify
various approximate effective models used by physicists by
deriving them as limiting cases of more fundamental theories. This
paper is devoted to a class of such models that one sometimes
calls quantum Langevin dynamics. We show that quantum Langevin
dynamics arise naturally as the limit of a dynamics of a small
system weakly interacting with a reservoir where not only the
small system, but also  the reservoir is taken into account.
 We will call this version of a weak coupling limit
 the {\em extended weak coupling limit}, to
differentiate it from the better known {\em reduced weak coupling limit}, which
involves only the dynamics reduced to the small system.

 To our knowledge, the main idea of  extended weak coupling limit
 first appeared in the
 literature in the work of Accardi, Frigerio and Lu in
\cite{accardifrigerio} under the name of  {\em stochastic limit}.

Our approach is inspired by their work, nevertheless we think that
it is both simpler and more powerful.

The reader may also find it useful to compare the present work
with our previous paper \cite{derezinskideroeck1}, which describes
the extended weak coupling limit on a relatively simple (and less
physical) example of the {\em Friedrichs
  model}. \cite{derezinskideroeck1}, apart from presenting results, which we believe are
mathematically interesting in their own right, can be viewed as
a preparatory exercise for the present work.

\subsection{Quantum Markov semigroups}
Before we discuss  quantum Langevin dynamics,  we should recall a
better known class of effective dynamics --  that of {\em quantum
Markov semigroups} (or, in other words, completely positive unity
preserving time-continuous semigroups). They are often used as a
phenomenological description of quantum systems.
 It is well known that every  quantum Markov semigroup on
 $B(\caK)$, where $\caK$ is a finite dimensional Hilbert space,
 can be written  as $\e^{tL}$ where $L$ can be written
 in the so-called
 {\em Lindblad form} \cite{lindblad}
\begin{equation}\label{lindi}
L(S)= -\iu (\Upsilon S - S \Upsilon^*) +  \nu^* S \nu,   \qquad S \in
\caB(\caK),\end{equation}
 $\nu$ is an operator from $\caK$ to $\caK\otimes\frh$ for some auxiliary
Hilbert space $\frh$ and $\Upsilon$ is an operator on $\caK$ satisfying
\begin{equation}\label{condition on Ga1}
-\iu \Upsilon + \iu \Upsilon^*= -  \nu^* \nu.
\end{equation}
Note that
 given $L$, the operators $\Upsilon$ and $\nu$ are not defined uniquely.

\subsection{Reduced weak coupling limit}
\label{Reduced weak coupling limit}
It is generally assumed that only reversible (unitary) dynamics appear in
fundamental quantum physics. Nevertheless, in
phenomenological approaches researchers often apply
non-unitary quantum Markov semigroups to describe irreversible
phenomena.
 A possible
justification for their use is provided by the so-called weak
coupling limit, an idea that goes back to Pauli and van Hove
\cite{vanhove},
 and was made rigorous in an
  elegant work of E.~B.~Davies \cite{davies1}. Davies proved that if a small
quantum system is weakly coupled to the environment, then the
reduced dynamics in the interaction picture,  after rescaling the
time as $\lambda^{-2}t$, converges to a quantum Markov semigroup
defined on the observables of the small system.

To be more specific, consider a system given by a Hilbert space
$\caH:=\caK\otimes\Gamma_\s(\caH_\res)$, where $\caK$ is a finite dimensional
Hilbert space, $\caH_\res$ is the 1-particle space of the reservoir and
$\Gamma_\s(\caH_\res)$ is the corresponding bosonic Fock space. The composite
system is described by the dynamics generated by the self-adjoint operator
\begin{eqnarray}
H_\lambda&=&K\otimes1+1\otimes \d\Gamma(H_\res)
+\lambda( a^*(V)+ a(V)).
\label{paulif}\end{eqnarray}Here $K$ describes the Hamiltonian of the small
system, $\d\Gamma(H_\res)$ describes the dynamics of
the reservoir expressed by the second quantization of a self-adjoint operator
$H_\res$ on $\caH_\res$,
and $a^*(V)$/$a(V)$
describe the interaction between the small system and the reservoir, which we
assume to be given by  the creation/annihilation operators
of an operator $V\in\caB(\caK,\caK\otimes\caH_\res)$.

The notation that we use to define $H_\lambda$ is
  explained only
in Section \ref{sec: notation}
 and may be unfamiliar to some of the readers. Therefore
  let us describe the operators  appearing in (\ref{paulif}) with perhaps a
  better known (although less compact) notation.
To this end it is convenient to identify $\caH_\res$ with $L^2(\Xi,\d \xi)$,
 for
  some measure space $(\Xi,\d \xi)$, so
  that one can introduce  $a_\xi^*/a_\xi$ -- the usual creation/annihilation
  operators  describing
bosonic excitations of the reservoir. Let $H_\res$ be the multiplication
operator by a real
function $\Xi\ni \xi\mapsto x(\xi)$ and let $\Xi\ni \xi\mapsto v(\xi)\in
 \caB(\caK)$ be
the function describing the operator $V$. Then we have an alternative notation
\begin{eqnarray*}
\d\Gamma(H_\res)&=&
\int x(\xi)a_\xi^*a_\xi\d \xi,\\ a^*(V)&=&\int v(\xi) a_\xi^*\d
\xi,\\ a(V)&=&\int v^*(\xi) a_\xi\d x\i.\end{eqnarray*}

Operators of the form (\ref{paulif})
are often used in quantum physics  in  phenomenological descriptions
of a small quantum systems interacting with an environment.
Some varieties of (\ref{paulif})
are known
under such  names as the spin-boson, Fr\"ohlich, Nelson
 and polaron Hamiltonian.
 Following \cite{derzinski1}, we will call operators of the form (\ref{paulif})
{\em  Pauli-Fierz operators}. (Note, however, that some authors
use this name in a slightly different meaning).

The vacuum vector in $\Gamma_\s(\caH_\res)$ will be  denoted by
 $\Omega$.
Let $I_\caK:\caK\to\caH$ denote the isometric embedding, which maps a vector
$\phi\in\caK$ on $\phi\otimes\Omega\in\caH$. Note that $I_\caK^*:\caH\to\caK$
 equals $1_\caK{\otimes}\langle\Omega|$, and $I_\caK^*\cdot I_\caK$ is the
 conditional expectation from $\caB(\caH)$ onto $\caB(\caK)$.

One  version of the result of Davies says that under some mild assumptions
the following limit exists
\begin{eqnarray}
\Lambda_t(S)&:=&\lim_{\lambda\searrow0}
\e^{\i\lambda^{-2} tK}I_\caK^*\e^{-\iu t\lambda^{-2} H_\lambda}
S\otimes1 \ \e^{\iu t\lambda^{-2} H_\lambda}
 I_\caK\e^{-\iu\lambda^{-2} tK}\nonumber\\
&=&\lim_{\lambda\searrow0}
I_\caK^*\e^{\iu\lambda^{-2} tH_0}\e^{-\iu t\lambda^{-2} H_\lambda}
S\otimes1 \ \e^{\iu t\lambda^{-2} H_\lambda}
\e^{-\iu\lambda^{-2} tH_0} I_\caK
,\label{davi}\end{eqnarray}
and $\Lambda_t$ is a quantum Markov semigroup.
Thus we obtain a, possibly irreversible, quantum Markov semigroup as a limit
of a family of
reversible, physically realistic dynamics.
We also obtain a concrete expression for the generator of $\Lambda_t$.
More precisely,  $\Upsilon$ and $\nu$ appearing in  (\ref{lindi}) are uniquely
defined  in terms of
$K,$ $H_\res$ and $V$.


In the literature on both the reduced and the extended weak
coupling limit, one usually considers a nontrivial reference state
for the reservoir, whereas
 we reduce our treatment to a vector state.
This is justified since one can always represent the
reservoir state as a vector
 state via the GNS construction. In particular,
 in the case of a thermal  bosonic state, we can use
 the Araki-Woods representations of the CCR, so that the reservoir state is
 given by the Fock vacuum.
The free
 reservoir Hamiltonian and the interaction are modified appropriately.
 For this reason, it is not
always appropriate to call  (\ref{paulif}) a ``Hamiltonian''.
 In typical applications that we have in mind, the environment is a collection
of heat baths at various positive temperatures, and then it is natural  to
take $\d\Gamma(H_\res)$ to be the sum of their {\em Liouvilleans}.
 In this case,
$H_\lambda$ is not bounded from below, and  it probably should not
be called a ``Hamiltonian''.  On the other hand, the name
``Liouvillean'' is not appropriate either, since on the small
system $K$ is actually  the Hamiltonian, not the Liouvillean.
Following the terminology introduced in \cite{derzinski1},  in
such a case $H_\lambda$ should be called a {\em semi-Liouvillean}.

\subsection{Quantum Langevin dynamics}

It is well known that a 1-parameter semigroup of contractions on a Hilbert
space can be written as a compression of a unitary group. This unitary group
is called a {\em dilation} of the semigroup.

 A similar fact
is true in the case of a quantum Markov semigroup. It has been noticed that
every such a semigroup can be written as
\beq\Lambda_t(S)=I_\caK^*\e^{-\iu tZ}\ S\otimes 1\ \e^{\iu tZ}I_\caK,\ \
S\in\caB(\caK).
\label{dila}\eeq
Here, $Z$ is a self-adjoint operator on a Hilbert space
$\caZ=\caK{\otimes}\Gamma_\s(\caZ_\res)$ for some 1-particle space $\caZ_\res$
and $I_\caK:\caK\to\caZ$ is defined analogously as before.

Unfortunately, in the literature there seems to be no consistent and uniform
terminology for this dilation. A possible name for the unitary dynamics
$\e^{\i t Z}$ seems to be a {\em Langevin-Schr\"odinger dynamics} or
a {\em stochastic Schr\"odinger dynamics} for the semigroup $\Lambda_t$.
 The corresponding dynamics in the Heisenberg picture, that is
$\e^{-\i t Z}\cdot\e^{\i tZ}$, will be called
a {\em quantum Langevin dynamics} or a {\em quantum stochastic dynamics}
 for the semigroup $\Lambda_t$.

The first construction of a quantum Langevin dynamics was probably
given by Hudson and Parthasaraty. In \cite{hudsonparathasaraty}
they
 introduced the so called
 \wdr{\emph{quantum stochastic differential equation}} - a generalization of the usual
 stochastic differential equation known from the Ito calculus. The
group $\e^{-\iu tZ}$ is
then given by the solution to this equation.

If the operators $\Upsilon$ and $\nu$ that appear in the
 generator of $\Lambda_t$ written in the Lindblad form
(\ref{lindi}) are given,  then there exists  a canonical
construction of the space $\caZ$ and of a Langevin-Schr\"odinger
dynamics $\e^{\iu tZ}$ on $\caZ$, which apart from (\ref{dila})
satisfies the condition \beq\e^{-\iu
  t\Upsilon}=I_\caK^*\e^{-\iu t Z}I_\caK.\label{dilaa}\eeq
Thus $\e^{-\iu tZ}$ is a dilation of the contractive semigroup $\e^{-\iu
 t\Upsilon}$ and $\e^{-\iu tZ}\cdot \e^{\iu tZ}$ is a dilation of the quantum
 Markov semigroup $\e^{tL}$.

In this construction, at least formally, $Z$ can be written in the form of a
  Pauli-Fierz operator
\beq \label{formally}Z=\frac12(\Upsilon+\Upsilon^*)  + \d \Gamma
(Z_\res )
 +\wdr{\frac{1}{\sqrt{2\pi}}} a(|1\rangle \otimes\nu) + \wdr{\frac{1}{\sqrt{2\pi}}} a^*(|1\rangle \otimes\nu).\eeq
The interaction that appears in (\ref{formally}) is quite singular
and
  difficult mathematically.
It is an example of a so-called {\em quantum white noise}
\cite{attalreview}.

The equation (\ref{dila}) suggests that  quantum Langevin dynamics have
 perhaps more physical
content than being just a mathematical device, and could be used
as effective dynamics describing a small system interacting with
environment. In fact,
 physicists (see e.g.\  \cite{gardinerzoller}) often use such quantum
Langevin dynamics
 to describe the interaction of  a small system with
 an environment, e.g.\ with several heat baths.

Quantum Langevin dynamics are also often used to describe
 processes involving ``continuous quantum measurements'' \cite{barchielli}.
 One can then introduce observables
 describing ``measurements performed in a given interval of time''.
Observables corresponding
 to measurements in non-overlapping time intervals commute,
 which can be a reasonable
 assumption in some idealized situations.

Note that the generator of a Langevin-Schr\"odinger dynamics
 is necessarily unbounded from
below. This is often put forward as an argument against physical
relevance of quantum Langevin dynamics.
 This  argument is actually not  justified, since unbounded from below
generators of dynamics appear naturally in physics, especially in
positive temperatures. We have seen such a situation when we
discussed (\ref{paulif}), since semi-Liouvilleans are typically
unbounded from below. (See also a remark at the end of
 Subsection \ref{Reduced weak coupling limit}).

\subsection{Extended weak coupling limit}

In \cite{accardifrigerio}, it was proposed by Accardi et al. that
one could extend the idea of the weak coupling limit from the
reduced dynamics to the dynamics on the whole system, and as a
result one can
 obtain a justification of using quantum Langevin dynamics to describe
quantum systems. They called their version of the weak coupling
limit the {\em stochastic limit}. In our opinion, this name is not
the best chosen, since the reduced weak coupling limit is just as
``stochastic'' as the extended one. Therefore we will use the
name {\em extended weak coupling limit}.

The reduced weak coupling limit in the form considered by Davies
 has a rather clean
mathematical formulation. Therefore, it was quickly appreciated by the
mathematical physics community. The extended weak coupling limit
 is inevitably
somewhat more
complicated, in particular since
 it involves constructions that are, to a certain extent,
arbitrary. Nevertheless, we believe that the idea of the extended weak coupling
 limit   is
valuable and sheds light on models used in physics, especially in quantum
optics and quantum measurement theory.
 In our paper we would like to state and prove
 a new version of the  extended weak coupling limit.

We start again from a dynamics generated by a ``Pauli-Fierz
operator'' (\ref{paulif}). As we discussed above, the reduced weak
coupling limit leads to a quantum Markov semigroup with the
generator given in a Lindblad form involving the operators
$\Upsilon$ and $\nu$.  Given these data, we have a canonical construction of a
quantum  Langevin-Schr\"odinger dynamics
 $\e^{-\iu tZ}$ acting on the ``asymptotic space'' $\caZ$ such that
(\ref{dila}) and (\ref{dilaa}) are
 satisfied.  We also
construct an appropriate identification operator $\Gamma(J_\lambda)$, which is
a partial isometry
mapping the physical space $\caH$
into the asymptotic space $\caZ$. Its main role is
to scale the physical energy. There is some arbitrariness in the construction
of the identification operator, since the frequencies away from the Bohr
frequencies (differences of eigenvalues of $K$) do not matter in the limit
$\lambda\searrow0$. Finally, one needs what we call the ``renormalizing
operator'' $Z_\ren$, which takes care of the trivial part of the dynamics
involving the
  eigenvalues of $K$. The main result of our paper can be stated as
\begin{eqnarray}
\s^*-\lim_{\lambda\searrow0}\e^{\iu\lambda^{-2}
 tZ_\ren}\Gamma(J_\lambda)\e^{-\iu
   \lambda^{-2} tH_\lambda}\Gamma(J_\lambda)^*&=&\e^{-\iu tZ},
\end{eqnarray}
where $\s^*-\lim$ denotes the strong* limit.
Thus $\e^{-\iu t(Z+\lambda^{-2}Z_\ren)}$ can be viewed as the effective dynamics
in the limit of $\lambda\searrow0$.

Note that in the Heisenberg picture we obtain for any $B\in\caB(\caZ)$
\begin{eqnarray*}
&&\s^*-\lim_{\lambda\searrow0}\e^{\iu tZ_\ren}
\Gamma(J_\lambda^*)\e^{-\iu
   \lambda^{-2} tH_\lambda}\Gamma(J_\lambda)B\Gamma(J_\lambda^*)\e^{\iu
   t\lambda^{-2}H_\lambda}\Gamma(J)^*\e^{-\iu tZ_\ren}\\&=&
\e^{-\iu tZ} B\e^{\iu tZ}.
\end{eqnarray*}
Replacing $B$ with $S\otimes1$,
 pretending $J_\lambda$ is unitary (which is justified, see e.g.\ Remark \ref{rem: J unitary} or expression \eqref{eq: almost unitarity}),
taking the conditional expectation
 $I_\caK^*\cdot I_\caK$
 of both sides and using (\ref{dila})
we retrieve (\ref{davi}) -- the reduced weak coupling limit.

One can also choose $B$ of the form $1 \otimes A$ such that the strong limit $ \Gamma(J_\lambda)B\Gamma(J_\lambda^*)$  as $\la \searrow 0$, exists. In that case, one can study fluctuations of reservoir quantities, see Theorem \ref{thm: algebra}.

We can summarize the results of our paper in the following diagram
(w.c.l. stands for weak coupling limit):
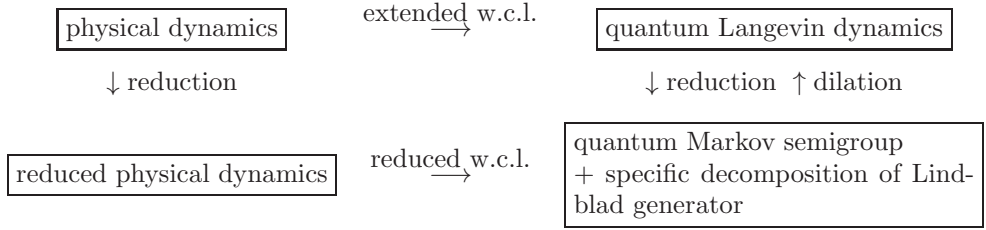
\[\begin{array}{ccc}
\hbox{\fbox{physical
    dynamics}}&\mathop{\longrightarrow}\limits^{\hbox{extended w.c.l.}}
&\hbox{\fbox{quantum Langevin dynamics}}\\[3mm]
\downarrow\hbox{reduction}&&\downarrow\hbox{reduction}\
\uparrow\hbox{dilation} \\[3mm]
\hbox{\fbox{reduced physical dynamics}}
&\mathop{\longrightarrow}\limits^{\hbox{reduced w.c.l.}}
&\hbox{\fbox{\parbox{150pt}{ quantum Markov semigroup \\+ specific
decomposition of Lindblad generator}}}
\end{array}\]

\subsection{Comparison with previous results}
As mentioned already, we are surely not the first to come up with
the concept of the extended weak coupling limit. Although the
original idea is attributed to Spohn \cite{spohnkinetic}, the
field was pioneered by Accardi et al.\ in \cite{accardifrigerio}
and a long list of works on the subject can be found in the book
\cite{accardibook}.   Recently, an interesting generalization has
been made by \cite{gough05}.

On the heuristic level, the ideas of the extended weak coupling
limit have been expressed by some physicists, e.g. by Gardiner and
Collett, see \cite{gardinercolletmaster} and Section 2.5 of
\cite{barchielli}.

The same idea was also applied to the low-density limit in
\cite{rudnickialicki} and \cite{accardilulowdensity}, see also
 \cite{pechen}. (The ``reduced low
density limit ''has been put on rigorous footing in
\cite{dumckelow})

 Most previous results we are aware of, have the
following form:
 For a Hilbert space $\caR$, let $\Phi(f)\in \Ga_{\mathrm{s}} (\caR)$ be the
 exponential vector for the 1-particle vector
 $f\in\caR$: \beq \Phi(f)= \exp\left(a^*(f)\right)\Omega.\eeq
Let $u,v \in
\caK, f,g \in \caH_\res, s_1 <t_1,s_2 <t_2 \in \bbR$ and put
$W_t^\la:= \e^{\i \la^{-2} t H_0} \e^{-\i \la^{-2} t H_\la}$.
Then, with all symbols having the same meaning as in the
introduction above,
\begin{eqnarray}\label{thm: conv accardi}
& \left \langle  u \otimes \Phi\left(\la
\mathop{\int}\limits_{s_1/\la^2}^{t_1/\la^2} \e^{-\i u H_\res} f
\d u \right)\Big| (W_t^\la)^* (S \otimes 1) W_t^\la
\, v \otimes \Phi \left(\la \mathop{\int}\limits_{s_2/\la^2}^{t_2/\la^2}
  \e^{-\i u H_\res} g \d u \right)   \right \rangle   & \nonumber \\
&\mathop{\rightarrow}\limits_{\la \to 0} \left \langle  u \otimes
\Phi (1_{[s_1,t_1] } \otimes f ) \Big| W_{t}^* (S \otimes 1) W_{t}
\, v \otimes \Phi (1_{[s_2,t_2] } \otimes g )  \right \rangle &
\end{eqnarray}
where $ W_t$ is the solution of an appropriate
Langevin Schr\"odinger differential equation on the space $\caK
\otimes \Ga_\s (L^{2}(\bbR) \otimes \caH_\res)$ and
$1_{[\cdot,\cdot]}$ is the indicator function of the interval
$[\cdot,\cdot]$.

%

Note that both our approach and (\ref{thm: conv accardi}) express essentially
the same physical idea. The scaling that we use to define $J_\lambda$ is
implicit in (\ref{thm: conv accardi}).
The main advantages of our approach  with respect to the previous
works are \ben
\item{The asymptotic space  $\caK
\otimes \Ga_\s (L^{2}(\bbR) \otimes \caH_\res)$ considered in
 (\ref{thm: conv accardi})  is much larger than the asymptotic space that we
use (which is introduced in Subsect \ref{sec: asymptotic}). One can argue that
our choice is  more
natural and ``tailor-made'' for
the problem at hand -- it closely resembles the original physical space
without introducing  unnecessary degrees of freedom.}
\item{ We prove convergence in the $*$-strong sense, instead of (as outlined
above) convergence of matrix elements of a class of rescaled coherent
vectors. This is mathematically cleaner and more flexible.}
\item{Our approach allows to consider also limits of certain
reservoir observables, see in particular Theorem \ref{thm:
algebra}}.
\item{We highlight the clear connection  between the work of Davies \cite{davies1} and D{\"{u}}mcke \cite{duemcke}, and extended weak coupling limits.
The latter follows rather easily from the results in
\cite{davies1} and \cite{duemcke}}.
 \een

A less important point of difference is the following:
In the early works on the weak coupling limit,   quasifree  reservoirs were
fermionic. If one chooses bosonic reservoirs, as
we do, one has to control the unboundedness of the interaction term (since the
bosonic creation and annihilation operators are unbounded). Although this is
not difficult, see Theorem \ref{thm: existence
hamiltonian}, we know of no place in the literature on the weak coupling limit
 where this difficulty is addressed. Of course, it is possible (and easy)
 to describe a version of
our result where the Hamiltonian $H_{\res}$ is fermionic.

From the physical point of view, our results justify a lot of the
manipulations one does with quantum Langevin dynamics (this is
discussed in detail in \cite{derezinskideroekmaes}). In
particular, Theorem \ref{thm: algebra} allows to identify
fluctuations of reservoir number operators with limits of
reservoir observables. These reservoir number operators (more
specifically: their fluctuations) are heavily studied objects, see
e.g.\ \cite{barchielli,boutenkummerer,deroeckmaesfluct}.
\subsection{Outline}

In Section \ref{sec: dilation}, we construct  a Langevin-Schr\"odinger
 dynamics
associated with a specific decomposition of a Lindblad generator.
 In the first subsection of Section \ref{sec: assumptions} we
introduce the class of our physical models considered in our paper
-- Pauli-Fierz operators.
 In the remaining subsections of Section \ref{sec: assumptions}
 we describe how to connect the setup of
the physical model with that of the corresponding
 quantum Langevin dynamics.
  Our results are listed in
Section \ref{sec: results} and their proofs are postponed to Section
\ref{sec: proofs}.

\noindent{\bf Acknowledgments.}
The research of J.~D. was  partly supported by the
Postdoctoral Training Program HPRN-CT-2002-0277 and the Polish KBN grants
 SPUB127 and 2 P03A 027 25. Part of the work was done  when both authors
 visited the
 Erwin Schr\"odinger Institute (J.~D. as a Senior Research Fellow), as well as
 during a visit of
J.~D.  at  K.~U.~Leuven supported by a grant of the ESF. W.~D.~R.
is a Postdoctoral Fellow supported by FWO-Flanders.

\section{Preliminaries and notation}\label{sec: notation}
 We will use the formalism of second quantization, following
 the conventions adopted in \cite{derzinski1}.

  For a Hilbert space $\caR$ and $n \in \bbN$, we recall
the projector $\Sym^n $, which projects elements of the tensor power
 $\otimes^n \caR
$ onto symmetric tensors. Its range will be denoted
$\Ga_{\s}^n(\caR)$ -- it is the $n$-particle subspace of the
bosonic Fock space over $\caR$.
 The symmetric (bosonic) second quantization of $\caR$ is hence
defined as \beq \sfock (\caR ) = \mathop{\oplus} \limits_{n
=0}^{\infty} \Ga_{\s}^n(\caR).\eeq

Note that we use the convention that $\otimes$ and $\oplus$ denote
the tensor product and the direct sum in the category of Hilbert
spaces. Sometimes we will use their algebraic counterparts.
If $\caD_1$ is subspace of a
Hilbert space $\caR$, then
 \begin{equation}
 \Gammal_{\s}^n
(\caD_1 ) = ({\otimesal}^n \caD_1)  \cap \Ga_{\s}^n(\caR),
\end{equation}
where $\otimesal$ denotes the algebraic tensor product. We
will often need
 \begin{equation}
 \Gammal_\s(\caD_1)
=  \Span \left\{ \psi \, | \, \psi \in \Gammal_{\s}^n(\caD_1), n \in
\bbN \right \}.
 \end{equation}

For $R \in \caB(\caK,\caK \otimes \caR)$, we heavily use the
generalized creation and annihilation operators $a(R)$ and
$a^*(R)$, as defined in \cite{derzinski1}. Actually, we need even
a slightly more general definition which is given now.

Assume that $\caD_1$ is a dense subspace of the Hilbert space
$\caR$ and
 $R^* : \caK
\otimesal \caD_1  \to  \caK   $ is an unbounded operator. Let $R$
stand for the adjoint of $R^*$ in the sense of quadratic forms.
(Note that the adjoint in the sense of forms is different from the
adjoint in the sense of operators.) Define for all $n \in \bbN$
\beq \label{def: general forms} a(R) \psi:= \sqrt{n} \left( R^*
\otimes \Sym^{n-1} \right)  \psi, \qquad \psi \in \caK \otimesal
 \Gammal_{\s}^n
(\caD_1 ). \eeq  $a(R)$ is well defined as an unbounded operator
and it defines a quadratic form on  $\caK \otimesal
\Gammal_{\s}(\caD_1) $. Denote by $a^*(R)$ its adjoint in the
sense of quadratic forms.

We write $\Om$ for the vacuum vector in $\sfock(\caR)$: \beq \Om =
1 \oplus  0 \oplus 0 \otimes 0 \oplus \ldots
 \eeq

$\mathrm{s}-\lim$ will denote the strong limit. We say that the
operators $A_{\la \in \bbR^+} \in \caB(\caR)$ converge
$*$-strongly to $A \in \caB(\caR)$ (notation: $\s^*-\lim_{\la
\downarrow 0} A_\la=A$) if \beq \s-\lim_{\la \downarrow 0} A_\la=A
\qquad \textrm{and} \qquad \s-\lim_{\la \downarrow 0} A^*_\la=A^*
\eeq

If $A$ is an operator, we will write
\[\Re A:=\frac12(A+A^*),\ \ \ \Im A:=\frac{1}{2\iu}(A-A^*).\]

Our typical Hilbert space will be the tensor product of two Hilbert spaces. We
will usually write $A$, $B$ for $A\otimes 1$ and $1\otimes B$.

\section{Dilations}
\label{sec: dilation}
\subsection{Unitary dilation of a contractive semigroup}

Let $\caK$ be a Hilbert space and let the family $\Theta_{t \in
\bbR^+}$ be a contractive semigroup on $\caK$:
\begin{equation}  \Theta_t \Theta_s =\Theta_{t+s}, \qquad \| \Theta_t \| \leq 1,    \qquad t,s
\in \bbR^+ .   \end{equation}
 \begin{definition}
 We say that $(\caZ,I_\caK,U_{t\in\bbR})$
is a unitary dilation
  of  $\Theta_{t \in \bbR^+}$ if
 \ben
  \item{$\caZ$ is a Hilbert space and $U_{t\in\bbR} \in \caB(\caZ)$ is a unitary one-parameter group; }
 \item{$\caK
\subset \caZ$ and $I_{\caK}$ is the  embedding of $\caK$ into $\caZ$;
}
\item{
 for
all $t \in \bbR^+$
\begin{equation} I_{\caK}^* {U}_t I_{\caK} =\Theta_t. \end{equation}} \een
\end{definition}

Assume that
 $\caK$ is finite-dimensional and the semigroup $\Theta_t$ continuous. Then there
 exists  a dissipative operator $- \iu \Upsilon \in \caB(\caK)$,   \beq
-\iu \Upsilon + \iu \Upsilon^* \leq 0,\eeq  such that $\Theta_t=
\e^{-\iu t \Upsilon}$.


\subsection{Quantum Langevin dynamics}

 Let the family ${\Lambda}_{t \in \bbR^+}$ be a  semigroup on $\caB (\caK)$:
\begin{equation}  {\Lambda}_t {\Lambda}_s ={\Lambda}_{t+s}
,  \qquad t,s
\in \bbR^+.    \end{equation}
\begin{definition}
 We say that $(\caZ,I_\caK,U_{t\in\bbR})$
is a   Langevin-Schr\"odinger dynamics for
  ${\Lambda}_{t \in \bbR^+}$ if
 \ben
 \item{$\tilde\caZ_\res$
is a Hilbert space and $U_{t\in\bbR} $ is a one-parameter
 unitary group on $\caZ:=\caK{\otimes}\tilde\caZ_\res$;}
 \item{$\Omega$ is a normalized vector in $\tilde\caZ_\res$ and
$I_\caK(\psi):=\psi{\otimes}\Omega\in \caZ$
 is the corresponding embedding of
$\caK$ into $\caZ$};
\item{
 for
all $t \in \bbR^+$ and all $S \in \caB(\caK)$,
\begin{equation} I_{\caK}^* {U}_{-t} S{\otimes}1 U_{t}I_{\caK} ={\Lambda}_t (S). \end{equation}} \een
\label{defi}\end{definition}

The Heisenberg dynamics $\e^{\i tZ}\cdot \e^{-\i tZ}$ corresponding to a
Langevin-Schr\"dinger dynamics will be called a {\em quantum Langevin
  dynamics}.

\begin{definition}
We say that $\Lambda_{t\in\bbR_+}$ is a quantum Markov semigroup iff it is a
semigroup on $\caB(\caK)$ such that for any $t\in\bbR_+$ the map $\Lambda_t$ is
completely positive and preserves the unity.
\end{definition}

Clearly, if a semigroup $\Lambda_t$ admits a Langevin-Schr\"odinger dynamics
  in the  sense of Definition \ref{defi},
 then it is a quantum Markov semigroup.

Again, assume that $\caK$ is finite dimensional.
Assume that $\Lambda_t$ is a continuous quantum Markov semigroup,
 so that we can define its generator $L$ and
we have $\Lambda_t=\e^{tL}$.
Recall that then there exists a
 dissipative operator $\Upsilon$ on $\caK$, another finite dimensional
Hilbert space
 $\frh$
 and an
operator $\nu \in \caB(\caK,\caK \otimes \frh)$, satisfying the
condition
\begin{equation}\label{condition on Ga}
-\iu \Upsilon + \iu \Upsilon^*= -  \nu^* \nu,
\end{equation}
such that
\begin{equation}
L(S)= -\iu (\Upsilon S - S \Upsilon^*) +  \nu^* S \nu,   \qquad S \in
\caB(\caK).
\label{lind}\end{equation}


\begin{vetremark}
If we choose an orthonormal basis $b_1,\dots,b_d$ in $\frh$, then
$\nu$ can be represented as a family of operators $\nu_1,\dots,\nu_d\in
\caB(\caK)$, and then (\ref{lind}) can  be rewritten as
\begin{equation}
L(S)= -\iu (\Upsilon S - S \Upsilon^*) +\sum_{j=1}^d   \nu_j^* S \nu_j,
 \qquad S \in
\caB(\caK).
\end{equation}
\end{vetremark}

\subsection{Construction of a Langevin-Schr\"odinger dynamics}
\label{sec: decomp of nu}

Let $\caK$, $\frh$ be finite dimensional Hilbert spaces, $\Re\Upsilon$ a
self-adjoint operator on $\caK$ and $\nu$ an operator from $\caK$ to
$\caK\otimes\frh$. Setting $\Im\Upsilon:=\nu^*\nu$ we obtain a dissipative operator
$\Upsilon:=\Re\Upsilon+\i\Im\Upsilon$ on $\caK$.

Given the data $(\caK, \Re \Upsilon,\frh,\nu)$ as above,
we will construct a dilation
 for $\e^{\i t\Upsilon}$, which at the
same time is a  Langevin-Schr\"dinger dynamics for $\e^{ tL}$.

Introduce the operator $Z_{\res}$ on $\caZ_\res:=L^2(\bbR) \otimes \frh \cong
L^2(\bbR,\frh)$ as the operator of multiplication by the variable
$x\in\bbR$:
\[
(Z_\res f)(x): =x   f(x).
\]
Put
\begin{equation}
\caZ =\caK \otimes\Gamma_\s( \caZ_\res).
\end{equation}

We define an unbounded linear functional  on $L^2(\bbR)$
 with domain
$L^1(\bbR) \cap L^2(\bbR) $, denoted $\langle 1|$, by the obvious
prescription
\[
\langle 1 | f  = \int_{\bbR} f.
\]
By $| 1 \rangle$,
 we denote the adjoint of $\langle 1 |$ in the
sense of forms.

We will also use the quadratic form
from $\caK$ to  $\caK \otimesal\left(
L^2(\bbR,\frh)\cap L^1(\bbR,\frh) \right) $:
 \beq \label{def: forms W}
 |1 \rangle \otimes \nu \eeq
%
%
%
Consider \beq \caD :=\caK \otimesal
\Gammal_{\s}\left(\Dom(Z_\res)\right), \eeq which is a  dense
subspace of $\caZ$.
%
%
%
As outlined in  Section \ref{sec: notation}, using the fact that
$L^2(\bbR,\frh)\cap
L^1(\bbR,\frh) \subset\Dom(Z_\res)$, we can
 define the
quadratic forms $a (|1 \rangle \otimes \nu )$ and $ a^*(|1 \rangle
\otimes \nu  )$ on $\caD$. Hence, also the following expressions
are quadratic forms on $\caD$:
 \begin{align}
\label{def: Zquad} {Z}^{+} &=\Upsilon + (2\pi)^{-\frac12}
a(|1 \rangle \otimes \nu) +
(2\pi)^{-\frac12}a^*(|1 \rangle \otimes \nu) + \d \Gamma
(Z_\res ), \\
{Z}^{-} & =\Upsilon^* + (2\pi)^{-\frac12}a(|1 \rangle \otimes \nu)
 + (2\pi)^{-\frac12}a^*(|1 \rangle \otimes \nu) + \d \Gamma (Z_\res ).
\end{align}
%

%
%
%
%
%
%
%
%
%
%
It will be convenient to
choose a family $b_{j \in \caJ} \in \frh$
and $C_{j \in \caJ} \in \caB(\caK)$ indexed by
 a finite index set $\caJ$ such that \beq \label{decomp of
nu} \nu= \sum_{j \in \caJ} C_j \otimes | b_j \rangle. \qquad \eeq
This can always be done, of course    in many ways. Define,
analogously to \eqref{def: general forms},
\beq a( \e^{\iu tZ_\res}|1\rangle \otimes b_j ), \qquad a^*(
\e^{\iu tZ_\res} |1\rangle  \otimes b_j ), \eeq as quadratic forms
on $\caD$. Note the equality \beq \label{eq: link gen creation} a
(|1 \rangle \otimes \nu) = \sum_{j \in \caJ} C^*_j \otimes
a(|1\rangle \otimes
 b_j). \eeq
%

For $a\leq b$, let $\triangle_n[a,b] \subset \bbR^n$ be the simplex
 \beq
\triangle_n[a,b]:=\left\{
(t_1,\dots,t_n)\ :\ a<t_1 < \ldots <
 t_n <b\right\}.
 \eeq

Set
\begin{eqnarray} C_{j}^+&=&C_j,\nonumber\\
C_{j}^-&=&C_j^*.\label{cj+}
\end{eqnarray}

Now we combine these objects into something that is a priori a
quadratic form,
 but
turns out to be a bounded operator. For $t\geq0$ we define
\begin{eqnarray}
U_t&:=&
\e^{-\iu t\d\Gamma(Z_\res)} \sum_{n=0}^\infty
\mathop{\int}\limits_{\triangle_n[0,t]}\d t_n\cdots\d t_1
\sum_{j_1,\dots,j_n\in
   \caJ}\ \sum_{\epsilon_1,\dots,\epsilon_n\in\{+,-\}}\nonumber \\
&&\times (-\iu)^n(2\pi)^{-\frac{n}{2}}\e^{-\i(t-t_n)\Upsilon}
C_{j_n}^{\epsilon_n}\e^{-\i(t_n-t_{n-1})\Upsilon}\cdots
C_{j_1}^{\epsilon_1}\e^{-\i(t_1-0)\Upsilon}\nonumber\\
&&\times\prod_{p=1,\dots,n:\  \ \epsilon_p=+} a^*(\e^{\i t_p
Z_R}|1\rangle\otimes b_{j_p}) \prod_{p'=1,\dots,n:\  \
\epsilon_{p'}=-}
a(\e^{\i t_{p'} Z_R}|1\rangle\otimes b_{j_{p'}});\nonumber\\
U_{-t}&:=&U_t^*.\label{def: group}
\end{eqnarray}
(In the above expression $\prod\limits_{p=1,\dots,n:\  \ \epsilon_p=+}$ should
be understood as the product over these indices $p=1,\dots,n$ that in addition
satisfy the condition $\epsilon_p=+$).

Finally, let $I_\caK$ be the  embedding
of $\caK \cong
\caK \otimes \Om $ into $\caK \otimes \sfock(\caZ_\res)$.
%
%
%
%
%
%

\begin{theorem}\label{group property}
Let ${Z}^{\pm}$ be as defined in (\ref{def: Zquad}) and $U_t$ as
defined in \eqref{def: group}.
\ben

\item{ The one-parameter
family of quadratic forms $U_t$ extends to a strongly continuous
unitary group on $\caZ$ and does not depend on the decomposition
(\ref{decomp of nu}).}

\item{
For $\psi,\psi' \in \caD$, the function $\bbR\ni t
\mapsto \langle \psi| {U}_t  \psi'
\rangle $ is differentiable away from $t=0$, its derivative
$t\mapsto \frac{\d}{\d t}
 \langle \psi| {U}_t  \psi' \rangle$ is continuous away from $0$ and
at
$t=0$ it has the left and the
right  limit equal respectively to
\begin{equation}
-\i\langle \psi| {Z}^+ \psi' \rangle  =
\lim_{t \downarrow 0}t^{-1}\langle \psi| ({U}_t-1)  \psi' \rangle,
\label{deri1}\end{equation}
\begin{equation}
-\i\langle \psi| {Z}^- \psi' \rangle
 =
\lim_{t \uparrow 0}t^{-1}\langle \psi| ({U}_t-1)  \psi' \rangle.
\label{deri2}\end{equation}
 }

\item{The triple $(\caZ,U_t,I_\caK)$ is a unitary dilation of the
semigroup $\e^{-\iu t \Upsilon}$ on $\caK$:
\begin{equation}\label{thm: group dil 1}
I_\caK ^*U_t  I_\caK = \e^{-\iu t \Upsilon}.
\end{equation}

}

\item{The triple
$(\caZ,U_t,1_\caK)$ is a   Langevin-Schr\"odinger dynamics for
the semigroup $\e^{ t L}$ on $\caB(\caK)$:
\begin{equation} \label{thm: group dil 2} I_\caK^* U_{-t} (S \otimes
1) U_{t} I_\caK =\e^{t L}(S), \qquad S \in \caB(\caK).
\end{equation}
}
 \een
\end{theorem}

We will say that $U_t$  constructed in the above theorem
is the Langevin-Schr\"odinger dynamics given by the
  data  $(\caK, \Re \Upsilon,\frh,\nu)$. Note that $U_t$ can be written as $\e^{-\i
  tZ}$ for a uniquely defined self-adjoint operator $Z$ on $\caZ$. Clearly,
  $\caD$ is not contained in the domain of $Z$ and the quadratic forms $Z^+$
  and $Z^-$ are not generated by the operator $Z$ (in fact, they are even not
  self-adjoint).   On an appropriate domain,
  $Z$ has the formal expression
\beq {Z} =\Re\Upsilon + (2\pi)^{-\frac12}a(|1 \rangle \otimes \nu)
 + (2\pi)^{-\frac12}a^*(|1 \rangle \otimes \nu) + \d \Gamma (Z_\res ),
 \eeq
which is the obvious ``self-adjoint compromise'' between $Z^-$ and
$Z^+$. This expression is formal since one needs a suitable
regularization to give it a precise meaning. Such a regularization, under an
additional assumption on the commutativity of the small system operators,
is discussed e.g.\ in \cite{chebotarev2}. See also
\cite{gregorattihamiltonian,vonwaldenfelssymmetric}.

\subsection{Alternative form of Langevin-Schr\"odinger equations}

Proofs of Theorem \ref{group property} are contained in the
literature, see e.g.\ \cite{maassen1}.  In any case, this theorem
involves well defined formulas and its proof follows by
straightforward computations, which  we leave  to the reader.
Nevertheless, we would like to mention a slightly different
(though equivalent) form of Langevin-Schr\"odinger dynamics, which is
closer to those usually appearing in the literature.

Let $\caG$ denote the normalized Fourier transform on $L^2(\bbR)$:
\[\caG f(s):=(2\pi)^{-\frac12}\int f(x)\e^{-\iu sx}\d x.\]
We can treat it as a unitary operator on $\caZ_\res$.
We  second
quantize $\caG$,
 obtaining an operator $\Gamma(\caG)$, which can be treated as an
operator on $\caZ$. Define also
\[(Tf)(x):=\iu\frac{\d}{\d x}f(x).\]
(A possible name for $T$ is the ``time operator''). Set \beq  \label{tilde}  \hat
Z_\res:= \caG Z_\res\caG^*,\ \ \ \hat T:= \caG
T\caG^*
 ,\ \ \ \hat U_t:=\Gamma(\caG)U_t\Gamma(\caG)^*  . \eeq
Note that \[\e^{-\iu t\d\Gamma(\hat Z_\res)}
=\Gamma\left(\exp\left(\scriptstyle -t\frac{\d}{\d
s}\right)\right).\] Then for $t\geq0$ the formula (\ref{def:
group}) transforms into
\begin{eqnarray}
\hat U_t&=&
\Gamma\left(\exp\left(\scriptstyle -t\frac{\d}{\d s}\right)\right)
\sum_{n=0}^\infty
\mathop{\int}\limits_{\triangle_n[0,t]}\d t_n\cdots\d t_1
\sum_{j_1,\dots,j_n\in
   \caJ}\ \sum_{\epsilon_1,\dots,\epsilon_n\in\{+,-\}}
  (-\iu)^n\nonumber \\
&&\times\e^{-\i(t-t_n)\Upsilon}
C_{j_n}^{\epsilon_n}\e^{-\i(t_n-t_{n-1})\Upsilon}\cdots
C_{j_1}^{\epsilon_1}\e^{-\i(t_1-0)\Upsilon}\nonumber\\
&&\times\prod_{k=1,\dots,n:\  \ \epsilon_k=+} a^*(\delta_{ t_k}
\otimes b_{j_k}) \prod_{k'=1,\dots,n:\  \
\epsilon_{k'}=-}
a(\delta_{t_{k'}}\otimes b_{j_{k'}}).\label{def: group1}
\end{eqnarray}
where $\delta_t$ denotes the deltafunction at $t\in\bbR$, and
 (\ref{def: group1})
 should be understood as a quadratic form between appropriate dense
  spaces.
 (\ref{def: group1}) is sometimes referred to in the literature
 as the representation by
\emph{integral
 kernels}. It was introduced by Maassen \cite{maassen1}. See also
\cite{vonwaldenfelsito,vonwaldenfelssymmetric,attalreview,goughasymptotic}
 and
 section VI 3.2 of \cite{meyer}.
Differentiating  (\ref{def: group1}) with respect to time we obtain
(at least
 formally)
\wdr{\begin{eqnarray} \iu\frac{\d}{\d t}\e^{\iu t\d\Gamma(\hat
Z_\res)}\hat U_t&=& \left(\Upsilon+a^*(\delta_t\otimes\nu)\right)
\e^{\iu t\d\Gamma(\hat Z_\res)}\hat U_t\nonumber\\
&&+\sum_{j \in \caJ}\nu_j^*\e^{\iu t\d\Gamma(\hat Z_\res)}\hat U_t
a(\delta_t\otimes b_j) ,\end{eqnarray}} which  essentially
coincides with what is
 known in the literature under the name of the  stochastic (or Langevin)
 Schr\"odinger
equation.

\section{The Pauli-Fierz operator}\label{sec: assumptions}
\subsection{Definitions and assumptions}
 Let $\caH=\caK \otimes \Gamma_\s (\caH_\res)$ where $\caK$, $\caH_\res$
 are Hilbert spaces. We assume that $\caK$ is  finite-dimensional.
 Fix a self-adjoint operator
 $H_{\res}$ on $\caH_\res$ and a self-adjoint operator $E$ on $\caK$.
 The operator $H_0$ on $\caH$  given as
 \[
 H_0=K + \d \Gamma (H_{\res})
 \]will be called the {\em free Pauli-Fierz operator}.
We choose a $V \in \caB(\caK,\caK \otimes \caH_\res)$ and we
recall the generalized creation and annihilation operators $ a(V)$
and $a^*(V)$ introduced in Section \ref{sec: notation}.
%
%
\begin{theorem} \label{thm: existence hamiltonian}
Set $H^I(t):=\e^{\i tH_0}(a^*(V)+a(V))\e^{-\i tH_0}$. Then
\begin{equation} \label{def: series for W}
W_{\la,t}\psi = \sum_{n=0}^{\infty}
\mathop{\int}\limits_{\triangle_n[0,t]} \d t_n\cdots \d t_1
\e^{\i t H_0}(\i\la)^n
H^I(t_n)\cdots H^I(t_1)\psi \end{equation} is
well defined for all $\psi\in  \caK \otimesal
\Gammal_{\s}(\caH_\res)$. $W_{\la,t}$ extends to a 1-parameter
unitary group on $\caK\otimes\Gamma_\s(\caH_\res)$ with
a self-adjoint generator $H_{\la}$. The finite particle space $\caK
\otimesal \Gammal (\caH_\res) $ belongs to the domain of $H_{\la}$
and on $\caK \otimesal \Gammal (\caH_\res) $,
\begin{equation} \label{pauli-fierz Hamiltonian}
 H_{\la}=H_{0}+\la \left( a(V) + a^* (V) \right).
 \end{equation}
\end{theorem}
$H_\lambda$ will be called the {\em full Pauli-Fierz operator}.

We write \beq K=\sum_{k \in \mathrm{sp}(K)} k 1_{\caK_k}, \eeq
where $k,1_{\caK_k}$, are the eigenvalues and the spectral projections
of $K$. We collect all Bohr frequencies in a set $\caF$:
\begin{equation} \label{def: set F}
\caF := \big \{\omega \in \bbR \big| \, \omega =k-k' \textrm{ for
some } k,k' \in \sp K \big \}.
\end{equation}
We again denote by $I_\caK$ the  embedding
of $ \caK = \caK \otimes
\Om$ into $ \caH $, where  $\Om \in \sfock(\caH_\res)$ is the
vacuum vector.
 %
 %
 %
%
%
%
%
%
%
%
%

We now list the assumptions that we will need in our
construction.\\

\begin{assumption}\label{ass: 1}
\textit{For any $\om\in\caF$ there exists a Hilbert space $\frh_\om$ and an open set
$I_\om\subset\bbR$ with $\om\in I_\om$ and an identification
\[\Ran 1_{I_\om}(H_{\res})\simeq L^2(I_\om)\otimes\frh_\om,\]
such that $H_{\res}$ is the multiplication by the variable $x\in I_\om$. We
assume that $I_\om$ are disjoint for distinct $\om\in\caF$ and we set
$I:=\cup_{\om\in\caF}I_\om$.
Thus if \beq
f\simeq \int\limits_I^\oplus f(x)\d x \in\Ran 1_I(H_{\res}),\label{ff}\eeq then
\[(H_{\res}f)(x)=xf(x),\]
for almost all $x$.}
\end{assumption}


%
%
%
%
%
%
%
%
%
\begin{assumption}\label{ass: 2}
\textit{For any $\om\in\caF$, there exists a measurable function \[I_\om\ni
  x\mapsto v(x)\in B(\caK,\caK\otimes\frh_\om)
\] such that for
 $f$  as in (\ref{ff}),
for almost all $x\in I$ we have
\[(Vf)(x)=v(x)f(x).\]}
 \textit{  Moreover, we assume that $v$ is continuous in $\caF$,
 so that for $\om \in \caF$ we can
unambiguously define $v(\om)\in\caB(\caK,\caK \otimes \frh_\om)$.
}
\end{assumption}

%
%
%
%
%


%
%
%
%
%
%
\begin{assumption}\label{ass: 4}
\textit{ For all }$S \in \caB(\caK)$,
\begin{equation}
  \int_{\bbR^+} \d t \,
 \| V^*  S{ \otimes}1
\ \e^{-\iu tH_0}   V \|  < \infty.
\end{equation}

\end{assumption}
%

%
%
%
%
%


%
%
%
%
%
%
%
%
\subsection{Asymptotic reduced dynamics}
\label{sec: asred}
 Let
\beq\frh := \mathop{\oplus} \limits_{\om \in \caF}  \frh_{\omega}.
\eeq

 We define the map $\nu_\om:\caK \to
\caK \otimes \frh_\om$
\[ \nu_\om :=\wdr{\sqrt{2\pi}} \sum_{\footnotesize{\left.\begin{array}{cc} k,k' \in
\sp K, \\ \om=k-k'\end{array}\right.} }
   1_{\caK_k}  v(\om)1_{\caK_{k'}},
\]
where $v(\om)$ is well-defined by Assumption \ref{ass: 2}. We also
define $\nu: \caK\to \caK \otimes \frh$
\[
\nu:=\sum_{\om \in \caF}
   \nu_\om
\]
 Under Assumption \ref{ass: 4}, we can define
\begin{eqnarray}\label{def: generator}
 \Upsilon &:= &-  \sum_{k \in \sp K } \iu
\int_0^\infty  1_{\caK_k}V^* \e^{-\iu t (K+H_\res-k)} V  \, 1_{\caK_k}  \d t\\
&=&
-\iu\sum_{\omega\in\caF}\ \sum_{k-k'=\omega}\int_0^\infty
 1_{\caK_k} V^*  1_{\caK_{k'}}\e^{-\iu t (H_\res
-\omega)} V  \, 1_{\caK_k} \d t.
 \end{eqnarray}
Remark that $- \iu \Upsilon$ is a dissipative operator and hence it
generates a contractive semigroup on $\caK$. Note that
\begin{eqnarray*}
\i\Upsilon-\i\Upsilon^*&=&
\sum_{\omega\in\caF}\ \sum_{k-k'=\omega}\int_{-\infty}^\infty
 1_{\caK_k} V^*  1_{\caK_{k'}}\e^{-\iu t (H_\res
-\omega)} V  \, 1_{\caK_k} \d t\\&=& \wdr{2\pi}\sum_{\omega\in\caF}\
\sum_{k-k'=\omega}
 1_{\caK_k} v^*(\omega)  1_{\caK_{k'}}v(\omega)  \, 1_{\caK_k}\ \ \  =\ \
 \nu^*\nu,\end{eqnarray*}
and thus  $\Upsilon$ and $\nu$  satisfy the condition
(\ref{condition on Ga}). Therefore,
\beq L(S)= -\iu (\Upsilon S - S \Upsilon^*) +  \nu^* S \nu,   \qquad S \in
\caB(\caK),\label{def: lind}\eeq
is the  generator of a quantum Markov semigroup.

\subsection{Asymptotic space and dynamics}\label{sec: asymptotic}
We introduce the asymptotic space and the asymptotic dynamics that we will use
 in our paper.
 The asymptotic reservoir one-particle spaces are
\begin{eqnarray}
\caZ_{\res_{\omega}} &:=&  L^2(\mathbb{R}, \mathfrak{h}_{\omega}
),
   \\
  \caZ_{\res} &:=&  \mathop{\oplus} \limits_{\om \in \caF} \caZ_{\res_{\omega}} = L^2(\bbR,\frh).
  \end{eqnarray}
For $\om \in \caF$, we have the ortogonal projections
\[
1_{\res_\om}:\caZ_\res  \to \caZ_{\res_\om}.\]

Let $Z_{\res}$ be the operator of multiplication by the variable
in $\bbR$ on $\caZ_{\res}$.

Clearly, we can construct from
$(\caZ,I_\caK,\nu,\Re \Upsilon)$ the Langevin-SChr\"odinger dynamics of Theorem \ref{group
property}. We denote it
by
${U}_{t}$ and its generator by ${Z}$.

Finally, we define a renormalizing Hamiltonian $Z_{\mathrm{ren}}$
on $\caZ$:
\begin{equation}
Z_{\mathrm{ren}} :={K} +
            \d \Gamma  ( \mathop{\oplus} \limits_{\om \in \caF}  \om
            1_{{\res}_\om} ).
\end{equation}

\subsection{Scaling}\label{sec: scaling}

For $\lambda > 0$, we define the family of partial isometries
$J_{\la,\om}:\caZ_{\res_\om}=L^2(\bbR,\frh_\om)
 \rightarrow  L^2(I_\om,\frh_\om)$, which on $g_\om \in
\caZ_{\res_\om}$ act as
\begin{equation}
  (J_{\la,\om}g_\om)(y)=   \left\{ \begin{array}{ll}  \frac{1}{\la}
        g_\om(\frac{y-\om}{\la^2}),  & \textrm{ if }   y\in I_\om; \\
        0,  & \textrm{ if } y\in\bbR\backslash I_\om. \\ \end{array}
  \right.
\end{equation}
Since $ L^2(I_\om,\frh_\om)\subset\caH_\res$, $J_{\la,\om}$ can be viewed as a
map from $\caZ_{\res,\om}$ to $\caH_\res$. We have
\begin{equation}
{J^*_{\la,\om} J_{\la,\om}}=1_{\lambda^{-2}(I_\omega-\omega)}(Z_\res)1_{\res_\omega}\mathop{\longrightarrow}\limits
_{\la \downarrow 0}^{\hbox{\tiny strongly}}
=1_{\res_\om } \qquad  J_{\la,\om} J_{\la,\om}^* = 1_{
I_\om}(H_{\res}).
\end{equation}
We set $J_\lambda:\caZ_\res\to\caH_\res$ defined for $g=(g_\om)_{\om\in\caF}$
by
\[J_\lambda g:= \sum_{\om\in\caF}
J_{\la,\om}g_\om .\]
Note that
\[J_\la J_\la^*=1_I(H_\res)
.\]

In what follows, we will mainly need the second quantized
$\Ga(J_\la)$, which will also be used to denote the
operator
\[
1 \otimes \Ga(J_\la) \in \caB(\caZ,\caH)
.\]

\begin{remark} \label{rem: J unitary} In the definition of
$J_\lambda$ there is a lot of
  freedom. What matters is what happens near the Bohr frequencies.
   \wdr{ In fact, essentially the only  requirement on $J_\la$ is that Lemma
  \ref{lem: conv of creation} holds and that both $J_\la^*J_\la$
  and $J_\la J^*_\la$ converge strongly to $1$. }
  \wdr{The form of $J_\la$ also reflects that different frequencies
  ``do not see each other" in the weak coupling limit (see e.g.\ \cite{frigeriogorini,accardigoughlu} for an
explicit discussion).}

%
%
\end{remark}

The following fact is immediate:

\begin{proposition}
We have
\begin{eqnarray*}
&& \label{thm: stochastic limit 3} \s^*-\lim_{\la \downarrow 0}
   \e^{\iu \la^{-2}t Z_{\mathrm{ren}}} \Ga  (J^*_{\la} )  \e^{-\i
 \la^{-2}t H_0 }   \Ga  (J_\la ) =\e^{- \iu t \d \Ga (Z_\res)}.
\end{eqnarray*}
\end{proposition}

\section{Results}\label{sec: results}

The full dynamics in the interaction picture will be denoted by
 \beq
   T_{\la} (t,t_0) = \e^{\iu  t H_0} \e^{-\iu (t-t_0
) H_\la} \e^{-\iu  t_0 H_0}.
 \eeq

We start with two  versions of older results by Davies
about the reduced weak coupling limit.
However, in most presentations of this subject
 contained in the
literature   the perturbation is
assumed to be bounded. This is not the case in Theorem
\ref{thm: davies}.

\begin{theorem}\label{thm: davies}
Assume  Assumptions \ref{ass: 1}, \ref{ass: 2}, \ref{ass: 4}. Let $T \leq
\infty$.
\ben\item Let
 $\Upsilon$ be as defined in
\eqref{def: generator}.
 Then
 \begin{equation}
 \lim_{\la \downarrow 0} I_\caK^*    T_{\la} (\la^{-2}t,\la^{-2}t_0)
   I_\caK= \e^{- \iu (t-t_0) \Upsilon }
 \end{equation}
 uniformly for  $T\geq t\geq t_0\geq -T$.
\item
 Let $L$ be as defined in
\eqref{def: lind}.
 Then
 \begin{equation}
 \lim_{\la \downarrow 0} I_\caK^*  T_{\la} (\la^{-2}t,\la^{-2}t_0)
 \  S{\otimes1}\
   T_{\la} (\la^{-2}t,\la^{-2}t_0)^* I_\caK= \e^{(t-t_0) L }(S)
 \end{equation}
 uniformly for  $T\geq t\geq t_0\geq -T$.
\een\end{theorem}

We will  prove Theorem \ref{thm: davies} 1) in Subsection
\ref{sec: proof of davies} --  it is an important step of the
proof of our main result. Theorem  \ref{thm: davies} 2) can be
proven by similar arguments, or, which is easier in our
framework, it follows immediately from Theorem \ref{thm:
algebra}.

%

%
The following result is a version of a result of D\"umcke
 \cite{duemcke}. Apart from its intrinsic interest,  we will need it as an
 important step in the proof of our main result.

\begin{theorem}\label{thm: duemcke}
Assume  Assumptions \ref{ass: 1}, \ref{ass: 2}, \ref{ass: 4} and
let $ T < \infty$, $\ell \in \bbN$ and
$S_1,\cdots,S_\ell\in\caB(\caK)$. Then
\begin{eqnarray}
&\mathop{\lim}\limits_{\la \downarrow 0} I_\caK^*
T_{\la}(\la^{-2}t,\la^{-2}t_\ell) S_\ell \cdots S_2
T_{\la}(\la^{-2}t_2,\la^{-2}t_1) S_1 T_{\la}(\la^{-2}t_1,\la^{-2}t_0)
 I_\caK & \nonumber\\
&=  \e^{- \iu (t-t_{\ell}) \Upsilon} S_\ell \ldots S_2\e^{ - \iu
(t_2-t_1) \Upsilon}
 S_1 \e^{ - \iu( t_1-t_0) \Upsilon} &
\end{eqnarray}
uniformly for
 ordered times $T\geq  t\geq
t_\ell \geq \ldots \geq t_1 \geq t_0\geq -T$.
\end{theorem}
Clearly, Theorem
 \ref{thm: davies}  1) is
a special case of Theorem \ref{thm:
duemcke}, corresponding to $\ell=0$, or all $S_i=1$.

\begin{remark}
Strictly speaking, Theorems \ref{thm: davies} 1) and \ref{thm:
duemcke} are somewhat different from  the results in
\cite{davies1} and \cite{duemcke}. In our setup, the latter are
consequences of Theorem \ref{thm: algebra} and Theorem \ref{group
property}. (See Remark \ref{rem: d and d as corollary}).
\end{remark}
%
%
Note that
the above results did not involve any dilations, nor the
identification operator $\Gamma(J_\lambda)$.

Our main result describes the extended weak coupling limit for Pauli-Fierz
 operators and reads
\begin{theorem}\label{thm: stochastic limit}
 Assume Assumptions \ref{ass: 1}, \ref{ass: 2},  \ref{ass: 4}.
Let $U_t$ be the Langevin-Schr\"odinger dynamics
constructed from $(\caZ,I_\caK, \nu,\Re \Upsilon)$ with $\caZ,\nu,\Upsilon$
defined in Section \ref{sec: asymptotic}. Let also
$Z_{\mathrm{ren}}$ be as defined in Section \ref{sec: asymptotic}.
Then,
\begin{eqnarray}
&& \label{thm: stochastic limit 2} \s^*-\lim_{\la \downarrow 0}
\Ga  (J^*_{\la} )T_\lambda(\lambda^{-2}t,\lambda^{-2}t_0)  \Ga (J_\la ) =
\e^{\iu t \d \Ga (Z_\res)} U_{t-t_0}\e^{-\iu t_0 \d \Ga (Z_\res)}
,   \\
&& \label{thm: stochastic limit 1} \s^*-\lim_{\la \downarrow 0}
\e^{\iu \la^{-2}t Z_{\mathrm{ren}}} \Ga  (J^*_{\la} ) \e^{-\iu
\la^{-2}t H_\la} \Ga  (J_\la ) =U_t.
\end{eqnarray}
\end{theorem}
\begin{remark}
Weaker versions of Theorem \ref{thm: davies} and Theorem \ref{thm:
duemcke} follow immediately from Theorem \ref{thm: stochastic
limit}. They are weaker because the uniformity in time is lacking.
\end{remark}

Remark that on $\Dom Z_\ren$, \beq \label{eq: zren and z
commute}[Z_\ren,U_t]=0\eeq as can be checked from the explicit
expression for $U_t$.
  The generator $Z_\ren$ could be considered as the free (i.e.\ $\caK$ and $\res$ are decoupled) Hamiltonian in the weak coupling
  limit and hence \eqref{eq: zren and z commute} expresses the
conservation of the 'decoupled' energy.
In the reduced weak coupling limit we have an analogous situation: the
  generator of the limiting quantum Markov semigroup $L$ commutes with the
  generator of the free evolution $\iu
  [K,\cdot]$.

A consequence of Theorem \ref{thm: stochastic limit} is now given.
Its advantage is that it does not involve explicitly the operators
$J_\la$.

Recall the notation in Assumption \ref{ass: 1}. Let $I \ni
x\mapsto g(x) \in \caB(\frh(x))$ be a measurable function such
that $\sup_{x \in I} \|g(x)\| < 1$ and $x \mapsto g(x)$ is
continuous in a
neighbourhood of $\caF$. Remark that this requirement makes sense
because of Assumption \ref{ass: 2}. Define the contractive
multiplication operator ${G} \in \caB(\caH_\res)$ as, \beq ({G}
f)(x) = g(x)f(x) \label{ggg}\eeq and remark that $\Gamma ({G} )$ is also a
contractive operator on $\Gamma_\s(\caH_\res)$. Let $ \caC$ be the
$C^*$-subalgebra of $\caB(\caH)$,
generated by \beq S \otimes 1 , \quad \textrm{and} \quad 1 \otimes
\Ga ({G} ),\eeq with $S \in \caB(\caK)$ and ${G}$ as defined above.
Let $\caC_{\mathrm{as}}$ be the $C^*$-subalgebra of $\caB(\caZ
)$ generated by
\beq S \otimes 1  \quad \textrm{and} \quad 1 \otimes \Gamma (1
\otimes p  )\eeq with $S \in \caB(\caK)$ and $p \in
\mathop{\oplus}\limits_{\om \in \caF} \caB(\frh_{\om})$.
%
%
%
\begin{proposition} There exists a unique $*$-homomorphism $\Theta : \caC \to
  \caC_{\mathrm{as}}$ such that
\beq \Theta ( S \otimes  \Ga ({G} )) = S \otimes
 \Gamma \left( 1 \otimes  \left( \mathop{\oplus} \limits_{\om \in \caF}
 g (\omega)
 \right)\right),\qquad S \in \caB(\caK),  \eeq where $G$ and $g$ are related
 by (\ref{ggg}).
We have
\beq
\label{connection Theta J} \Theta( A )=\s^*-\lim_{\la\searrow0} \Ga (J^*_{\la}
)  A \Ga  (J_\la ) ,\ \ \ A\in\caC.\eeq
\end{proposition}
\begin{theorem}\label{thm: algebra}
Assume Assumptions \ref{ass: 1} \ref{ass: 2}, \ref{ass: 4} and let
the family $U_t$ be as in Theorem \ref{thm: stochastic limit}. For
any $\ell \in \bbN$, any
$A_1,\ldots,A_{\ell}\in\caC$
 and (not necessarily ordered) times $t_0,t_1,\ldots,
t_\ell,t \in \bbR$,
\begin{eqnarray}
&\mathop{\lim} \limits_{\la \searrow 0}   I_\caK^*  T_\la (\la^{-2}t,
 \la^{-2} t_{\ell})         A_n \ldots A_2 T_\la (\la^{-2}t_2, \la^{-2}t_{1})   A_1   T_\la (\la^{-2}t_{1},\la^{-2}t_0) I_{\caK} & \nonumber\\
=&    I_\caK^*   U_{(t- t_{\ell})}     \Theta(A_n) \ldots \Theta(A_2)
U_{(t_2- t_1)}   \Theta(A_1)  U_{(t_1-t_0)} I_{\caK}.&
\end{eqnarray}


\end{theorem}
\begin{vetremark}\label{rem: d and d as corollary}
The results in \cite{davies1} and \cite{duemcke} correspond to
Theorem \ref{thm: algebra} where $A_1\ldots,A_{\ell}$ are elements
of $\caB(\caK)$ and hence
$\Theta(A_{1,\ldots,\ell})=A_{1,\ldots,\ell}$.
\end{vetremark}


%
%
%
%
%
%
%
%
%
%

\section{Proofs}\label{sec: proofs}

\subsection{Proof of Theorem \ref{thm: existence hamiltonian}\\ -- existence of
  the physical dynamics}

We will prove a somewhat stronger theorem.
Let $P_n$ be the orthogonal projector on $\caK \otimes \Ga_\s^n
(\caH_\res)$ and let the dense subspace $\caD_1$ od $\caH$ be defined as
\[ \psi \in \caD_1 \Leftrightarrow  \hbox{ there exists }C\hbox{ such that
  for }n=0,1,2\dots\hbox{  we have }
\|P_n\psi\|\leq\frac{C^n}{\sqrt{n!}}.
 \]
\begin{theorem} For $\psi\in\caD_1$, the series (\ref{def: series for W})
defining $W_{\lambda,t}\psi$ is absolutely convergent,
 belongs to $\caD_1$, is continuous wrt $t\in\bbR$ and we have
$W_{\lambda,t}
W_{\lambda,s}\psi
=W_{\lambda,t+s}\psi$, $\|W_{\lambda,t}\psi\|^2=\|\psi\|^2$.
Therefore, $W_{\lambda,t}$ extends uniquely to
 a strongly continuous unitary group on $\caH$. By
Stone's theorem, it has a self-adjoint generator $H_\lambda$,
and by a theorem of
Nelson,
 $\caD_1$ is a core for $H_\lambda$.
\end{theorem}

\proof It is enough to assume that $\lambda=1$, so that we will write
 $W_t$. We can also assume that $t\geq0$.
 Let $\psi\in\caD_1$ with $\|P_m\psi\|\leq
\frac{C^m}{\sqrt{m!}}$.
Note that we have
\begin{eqnarray*} \label{def: series for W1}
P_nW_{t}\psi& =& \sum_{q=0}^{\infty}
\mathop{\int}\limits_{\triangle_n[0,t]}\sum_{m=\max\{n-q,0\}}^\infty\d
t_m\cdots \d t_1 \\
&& \times \e^{\i t H_0}\i^n
P_nH^I(t_q)\cdots H^I(t_1)P_m\psi .\end{eqnarray*}
Note also that
\[\|H^I(t_q)\cdots H^I(t_1)P_m\|\leq\left\{\begin{array}{cc}
2^q\|V\|^q\frac{\sqrt{(m+q)!}}{\sqrt{m!}},&m\geq n-q,\\
0,&m<n-q.\end{array}\right.
\]
Therefore,
\begin{eqnarray*}
\|P_nW_{t}\psi\|&\leq&
 \sum_{q=0}^{\infty}
\sum_{m=\max\{n-q,0\}}^\infty
\frac{ (2t\|V\|)^q}{q!}\frac{\sqrt{(m+q)!}}{\sqrt{m!}}\frac{C^m}{\sqrt{m!}}
\\
&\leq&\sum_{p=n}^\infty\sum_{q=0}^p\frac{1}{\sqrt{p!}}
\frac{(2t\|V\|)^qC^{p-q}p!} {q!(p-q)!}\\
&\leq&\sum_{p=n}^\infty\frac{1}{\sqrt{p!}}(2t\|V\|+C)^p\\
&\leq&\frac{(2t\|V\|+C)^n}{\sqrt{n!}}\sum_{r=0}^\infty
\frac{(2t\|V\|+C)^r}{\sqrt{r!} }\\&=&
\frac{(2t\|V\|+C)^n}{\sqrt{n!}}C_1. \end{eqnarray*}
This proves the absolute convergence of the series and the fact that it
belongs to $\caD_1$. The rest of claims is now straightforward. \qed

\subsection{Decomposition of interaction}

\begin{lemma}\label{lem: existence integrable h}
There is a finite index set $\caJ$ and families $D_{j \in \caJ} \in
\caB(\caK)$ and $\phi_{j \in \caJ} \in \caH_\res$ such that
 \beq \label{def: decomp V} V= \sum_{j \in \caJ} D_j \otimes | \phi_j \rangle,
 \eeq
 and such that the function
 \beq  \label{eq: integrability of h} h(t) := \sum_{j,j' \in \caJ} |  \langle \phi_{j'} | \e^{-\iu t H_\res}
\phi_j \rangle |    \eeq is integrable: \beq \label{h integrable}
\|h \|_1:=\int_{\bbR} \d t \,  h (t)
 < +\infty. \eeq
Moreover, we can choose this decomposition so that
 all $\phi_{j \in
\caJ}$ are continuous in $\caF$  and for all  $j \in \caJ$
there is at most one  $\om \in \caF$
such that $\phi_j(\om)\neq0$. \end{lemma}

\proof Let $\{w_p\}_{p\in\caP}$ be an orthonormal
 basis of eigenvectors of $K$, so that $Kw_p=k_pw_p$. For each $p \in
 \caP$,
there exists a family $\{ \phi_{p,m} \}_{m \in \caM} $ in $ \caH_\res $ such
that
\[V w_p=\sum_{m \in \caM} w_m{\otimes}\phi_{m,p}.\]
We set $S=|w_m\rangle\langle w_m|$. Now
\begin{eqnarray*}
\langle w_p|V^*S\e^{\iu t H_0} V w_p\rangle
&=&\langle\phi_{p,m}|\e^{\iu t H_\res}\phi_{p,m}\rangle\e^{\iu tk_m}
\end{eqnarray*}
is integrable by assumption  \ref{ass: 4}.

Then we choose a partition of unity $\chi_\omega\in C_c^\infty(\bbR)$ together
with $\chi_\infty\in C^\infty(\bbR)$ such that $\chi_\omega=1$ on a
neighborhood of $\omega$, $\chi_\omega=0$ on a neighborhood of
$\caF\backslash\{\omega\}$ and $\chi_\infty=0$ on a neighborhood of $\caF$ and
$\sum_m\chi_m+\chi_\infty=1$. We
set $\phi_{m,p,\omega}:=\chi_\omega(H_\res)\phi_{m,p}$,
 $\phi_{m,p,\infty}:=\chi_\infty(H_\res)\phi_{m,p}$,
 $D_{m,p,\omega}=D_{m,p,\infty}:=
|w_m\rangle\langle w_p|$. Hence, the index set $\caJ$ is chosen
as $\caM \times \caP \times (\caF \cup \{\infty \} )$ and elementary properties of the
Fourier transform imply  the integrability of \eqref{eq:
integrability of h}. \qed

If for a given $j\in \caJ$ and $\omega\in\caF$,
 we have  $\phi_j(\omega) \neq 0$, then this $\om$
 will be referred to as $\om(j)$.
If for  a given $j$, there is no $\om\in \caF$ such that $\phi_j(\omega)
 \neq
 0$,  then $\om(j)$ is chosen arbitrarily.
For further reference let us record the identity \beq
v(\omega)=\sum_{j\in \caJ\ : \ \omega(j)=\omega}
 D_j\otimes|\phi_j(\omega)\rangle,\ \ \omega\in\caF.
\eeq

\subsection{Proof of Theorems \ref{thm: davies} \\ -- reduced weak coupling
 limit} \label{sec: proof of davies}

 For operators
$A_1,\dots ,A_p$ we will write
\[\prod_{i=1}^pA_i:=A_p\cdots A_1.\]
We will also write $D(t):=\e^{-\iu tK} D\e^{\iu tK}$.

Define \begin{eqnarray*}
&& G_{\la}(t,t_0) \\&:= &
\sum_{n=0}^{\infty} (\iu \la)^{-2n}
\int_{\triangle_{2n}[t_0,t]} \d t_1\ldots \d t_{2n}\\&&\times
 \prod_{i=1}^{n}
\,I_\caK^* H^{\mathrm{I}} (\la^{-2} t_{2i}) H^{\mathrm{I}}(\la^{-2}
t_{2i-1}) I_\caK \\
&=&
\sum_{n=0}^{\infty} (\iu \la)^{-2n}
\sum_{j_1,\dots,j_{2n}\in
\caJ} \int_{\triangle_{2n}[t_0,t]} \d t_1\ldots \d t_{2n}
\\&&\times\prod_{p=1}^n
D_{j_{2p}}^*(\lambda^{-2}t_{2p})D_{j_{2p-1}}(\lambda^{-2}t_{2p-1})
\langle\phi_{j_{2p}}|\e^{\iu\lambda^{-2}(t_{2p}-t_{2p-1})H_\res}
\phi_{j_{2p-1}}\rangle.\nonumber
\end{eqnarray*}

\begin{lemma}
For all $T \leq
\infty$, \beq \lim_{\la \downarrow 0}  \sup_{0 \leq t_0 \leq t \leq
T}\| I_\caK^* T_{\la}(\la^{-2}t,\la^{-2}t_0)
I_\caK- G_{\la}(t,t_0) \|=0. \eeq
\end{lemma}

\begin{proof}
Set
\begin{eqnarray} D_{j}^+&=&D_j,\nonumber\\
D_{j}^-&=&D_j^*.\label{dj+}
\end{eqnarray}
For $n=0,1,2,\dots$, let $\Pair(2n)$ denote the set of pairings of
$\{1,\dots,2n\}$. That means, $\sigma\in\Pair(2n)$ iff it is a permutation
 $\sigma\in
S_{2n}$ satisfying $\sigma(2p-1)<\sigma(2p)$, $p=1,\dots,n$,
 and $\sigma(2p-1)<\sigma(2p+1)$, $p=1,\dots,n-1$. We will write $\epsilon(p)=+$ for even $p$ and
$\epsilon(p)=-$ for odd $p$.  One can visualize the above
definitions as follows:
$\sigma(2p-1)$ corresponds to the $p$th creator in the order of increasing
time and $\sigma(2p)$ corresponds to the annihilator paired with this creator
by the Wick theorem.

Using first the Dyson expansion and then the Wick theorem we obtain
\begin{eqnarray}
&&I_\caK^*
T_\lambda(\lambda^{-2}t,\lambda^{-2}t_0)I_\caK\nonumber\\\label{nomu}
&=&\sum_{n=0}^\infty\sum_{\sigma\in\Pair(2n)}
\sum_{j_1,\dots,j_{2n}\in
  I} (\iu\lambda)^{-2n}
\int\limits_{\triangle_{2n}[t_0,t]}\d t_{2n}\cdots\d t_1\\&&
\times\prod_{i=1}^{2n}
 D_{j_{i}}^{\epsilon(\sigma(i))}(\lambda^{-2}t_{i})
\nonumber\\
&&\times\prod_{p=1}^n
\langle\phi_{j_{\sigma(2p)}}|\e^{\iu\lambda^{-2}
(t_{\sigma(2p)}-t_{\sigma(2p-1)})H_\res}
\phi_{j_{\sigma(2p-1)}}\rangle
\nonumber\\
&=:&\sum_{n=0}^{+\infty}C_n.\nonumber
\end{eqnarray}

Assume for simplicity $t_0=0$. Abbreviating $\| D\|:= \sup_{j \in \caJ} \| D_j \|$, we
obtain a uniform estimate
\begin{eqnarray}
&&\|C_n\|\nonumber\\
&\leq& (\|D\|\la^{-1})^{2n}
 \sum_{\pi \in \Pair(2n)}\int\limits_{\triangle_{2n}[0,t]} \d t_1
\ldots \d t_{2n}  \nonumber\\&& \times \prod_{p=1}^n
h(\la^{-2}(t_{\pi
(2p)}-t_{\pi (2p-1)}) ) \nonumber\\
&=&  \frac{ (\|D\|\la^{-1})^{2n}
}{2^n n!} \sum_{\pi \in S_{2n}}\int\limits_{\triangle_{2n}[0,t]} \d t_1
\ldots \d t_{2n}  \nonumber \\&& \times \prod_{p=1}^n h(\la^{-2}|t_{\pi
(2p)}-t_{\pi (2p-1)}| )  \nonumber\\
&=&  \frac{ (\|D\|\la^{-1})^{2n}
}{2^n n!(2n)!} \sum_{\pi \in S_{2n}}\nonumber\int\limits_{[0,t]^{2n}} \d t_1
\ldots \d t_{2n} \\&&\times
\prod_{p=1}^n h(\la^{-2}|t_{\pi
(2p)}-t_{\pi (2p-1)}| )  \nonumber\\
&\leq&  \frac{(\|D\|\la^{-1})^{2n}
t^n}{2^n n!} \left( \int_{-t}^{t} \d
s h(\la^{-2} |s| ) \right)^n \nonumber\\
&\leq&      \frac{(\|D \|)^{2n}}{2^nn!}
t^n \| h \|_1^n .  \label{eq: uniform estimate}
\end{eqnarray}
First we used that
each pairing can be represented by $2^nn!$ permutations. Then we allowed to
permute $t_1,\dots,t_{2n}$.
 The last
 inequality has been obtained by a change of integration
 variables.  The bound \eqref{eq: uniform estimate} shows that the
 series \eqref{nomu} is absolutely convergent. We will exploit
 this now since we estimate the series term by term.

 Given a pairing $\sigma$, the term in the sum (\ref{nomu})
 is estimated by
\beq\label{term} \la^{-2n}
\int_{\triangle_{2n}([0,t])} \d t_1 \ldots \d t_{2n} \,
\prod_{i=1}^n h(\la^{-2}(t_{\pi (2i)}-t_{\pi (2i-1)}) ). \eeq
We are going to show that (\ref{term})
 does not vanish
 only for the time  consecutive pairing: that is for the pairing given by
 the identity permutation (also called ``nonnested, noncrossing pairings" for obvious reasons).
Assume there is $i$ such that $\pi(2i)-\pi (2i-1)  >1$ and let $p$
be such that $ s_1:= t_{\pi (2i-1)}< t_{\pi(p)} <t_{\pi(2i)}=:s_2
$. Then
\begin{eqnarray}
&& \la^{-2n}\int_{\triangle_{2n}([0,t])} \d t_1\ldots \d t_{2n}
\prod_{i=1}^n  h(\la^{-2}(s_2-s_1) ) \nonumber\\
&\leq&    \la^{-2}  t^{n-2} (2\|h\|_1)^{n-1} \int^{t}_0  \d
t_{\pi(p)} \int^{t_{\pi(p)}}_0 \d s_1   \int_{t_{\pi(p)}}^t   \d
s_2 \,
  h( \la^{-2}(s_2-s_1)) \nonumber \\
&\leq&      t^{n-2} (2\|h\|_1)^{n-1} \int^{t}_0   \d t_{\pi(p)}
\int^{t_{\pi(p)}}_0 \d s_1 \int_{\la^{-2}(t_{\pi(p)}-s_1) }
^{\infty} \d u \, h(u).
\end{eqnarray}
The last line vanishes uniformly in $0 \leq t \leq T$ by the
dominated convergence theorem, since the expression is dominated
by $t^2 \|h \|_1 $, and the $\d u$-integral vanishes as $\la
\downarrow 0$ whenever $s_1<t_{\pi(p)}$. This ends the proof since
$G_{\la}(t,t_0)$ is the sum of all terms with time-consecutive
pairings.
\end{proof}%
Now notice that $G_\lambda(t,t_0)$ can be written in the form
familiar from the weak coupling limit for  Friedrichs
Hamiltonians. In fact, if we consider the Hilbert space
$\caK\oplus (\caK{\otimes} \caH_\res)$ with the Friedrichs-type
Hamiltonian
\[\tilde H_\lambda
:=\left[\begin{array}{cc}K&\lambda V\\\lambda V^*&K+H_\res\end{array}\right],\]
then we can write
\[G_\lambda(t,t_0)=\e^{\i \lambda^{-2}tK}I_\caK^*\e^{-\i\lambda^{-2}
 t\tilde H_\lambda}I_\caK
\e^{-\i \lambda^{-2}t_0K}.\] Therefore, we can  apply Theorem 2.1
in \cite{davies1}. More precisely, define
\begin{eqnarray} \label{def: K}
 Q_{\la,s} &:= &\la^{-2} \int_0^s \d u  \,  \e^{\iu \la^{-2} u K}
 I_\caK^* H^{\mathrm{I}} (\la^{-2} u)  H^{\mathrm{I}}(0)  I_\caK \nonumber \\
 &= & \int_0^{\la^{-2} s }  V^* \e^{-\iu u (K+H_\res)} V \d u
 \end{eqnarray}
 and remark that by Assumption \ref{ass: 4},
\ben
\item{For all $\tau_1>0$, there is $c>0$ such that  \beq \label{cond: davis1} 0 \leq s \leq \tau_1 ,\, \la \leq 1 \Rightarrow \| Q_{\la,s} \| \leq  c. \eeq}
\item{
For all $  0< \tau_0 \leq \tau_1 < \infty$, \beq \label{cond:
davis2} \lim_{\la \downarrow 0} \sup_{ \tau_0 \leq s \leq \tau_1}
\| Q_{\la,s}-Q \|=0,
 \eeq
with \beq Q :=  \int_0^{+\infty}   V^* \e^{-\iu u (K+H_\res)}V  \d
u < \infty .\eeq
 }
\een The aforementioned theorem by Davies allows us to conclude
from the above 1) and 2), and the fact that $\Upsilon = \sum_{k
\in \sp K} 1_{\caK_k} K 1_{\caK_k}$, that for all $T < \infty$,
\beq \label{eq: celebrated}\lim_{\la \downarrow 0} \sup_{ t_0 \leq
t \leq T } \| G_{\la}(t,t_0)- \e^{-\iu (t-t_0) \Upsilon} \| =0.
\eeq

\subsection{Proof of Theorem
\ref{thm: duemcke}\\ weak coupling limit for correlations}

We follow very closely the strategy of D\"umcke in \cite{duemcke}.
The case $\ell=0$ has been already proven.
For notational reasons, we restrict ourselves to the case  $\ell=1$.
 Higher $\ell$ are proven in exactly the same
way.

The theorem
for the case $\ell=1$ follows
immediately from Theorem \ref{thm: davies} and the following lemma:
\begin{lemma} For all $T \leq \infty $ and $S \in
\caB(\caK)$
 \begin{eqnarray}\nonumber
 &\lim\limits_{\la
\downarrow 0} \sup\limits_{-T\leq t_0< t'<t<T } &\| I_\caK^*
T_{\la}(\la^{-2}t,\la^{-2}t')\ S{\otimes} 1\  T_{\la}(\la^{-2}t',t_0)
 I_\caK\\&& - I_\caK ^*T_{\la}(\la^{-2}t,\la^{-2}t')
 I_\caK
S I_\caK^* T_{\la}(t',t_0) I_\caK \| =0.
  \label{eq: factorization} \end{eqnarray}
\end{lemma}

\begin{proof}
Using first the Dyson expansion and then the Wick theorem we obtain
\begin{eqnarray}
&&
 I_\caK^*
T_{\la}(\la^{-2}t,\la^{-2}t')\ S{\otimes} 1\  T_{\la}(\la^{-2}t',t_0)
 I_\caK\\&& - I_\caK ^*T_{\la}(\la^{-2}t,\la^{-2}t')
 I_\caK\nonumber\\
&=&\sum_{n=0}^\infty\sum_{\sigma\in\Pair(2n)}
\sum_{j_1,\dots,j_{2n}\in
\caJ} (\iu\lambda)^{-2n}\label{summa}\\
&&\times\sum_{p=0}^n
 \int\limits_{\triangle_{p}[t_0,t']\times
\triangle_{2n-p}[t',t]}\d t_{2n}\cdots\d t_1\nonumber\\&&
\times\prod_{i=p+1}^{2n}D_{j_{i}}^{\epsilon(\sigma(i))}(\lambda^{-2}t_{i})
S
\prod_{i'=1}^pD_{j_{i'}}^{\epsilon(\sigma(i'))}(\lambda^{-2}t_{i'})
\nonumber\\
&&\times
\prod_{p=1}^n
\langle\phi_{j_{\sigma(2p)}}|\e^{\iu\lambda^{-2}
(t_{\sigma(2p)}-t_{\sigma(2p-1)})H_\res}
\phi_{j_{\sigma(2p-1)}}\rangle
.\nonumber\\
&=:&\sum_{n=0}^{+\infty}C_n(S).\nonumber
\end{eqnarray}

Assume for simplicity $t_0=0$.
Exactly as in (\ref{eq: uniform estimate}), we prove that
\begin{eqnarray}
\|C_n(S)\|
&\leq&    \|S\|  \frac{(\|D \|)^{2n}}{2^nn!}
t^n \| h \|_1^n ,  \label{eq: uniform estimate1}
\end{eqnarray}
so we can again estimate the series \eqref{summa} term by
term.

The term in the sum (\ref{summa}) corresponding  to the pairing
$\sigma$ is estimated by \beq  (\frac{\|D \|}{\la})^{2n}\|S\|
\int_{\triangle_{p}[0,t']\times\triangle_{2n-p}[t',t]} \d t_1
\ldots \d t_{2n} \, \prod_{i=1}^n h(\la^{-2}(t_{\pi (2i)}-t_{\pi
(2i-1)}) ) .\eeq We are going to show all such terms with a pairing
crossing $t'$ vanish in the limit $\lambda\searrow0$.

 Assume
there is a $i$ such  that \beq  s_1:=
t_{\pi (2i-1)}< t' < t_{\pi(2i)} =:s_2.\eeq
 Then
\begin{eqnarray}
&& \la^{-2n}
 \int_{\triangle_{p}[0,t']\times\triangle_{n-p}[t',t]} \d t_1 \ldots \d
t_{2n} \,\prod_{i=1}^n \, h(\la^{-2}(s_2-s_1) )\nonumber \\
& \leq&  \la^{-2}  (2t\|h\|_1)^{n-1}    \int^{t'}_0 \d s_1
\int_{t'}^t
\d s_2  \,    h(\la^{-2}(s_2-s_1))  \nonumber \\
&\leq &  (2t\|h\|_1 )^{n-1}   \int^{t'}_0 \d s_1 \int_{\la^{-2}
(t'-s_1)}^{+\infty} \d u \,  h(u)  .\nonumber
\end{eqnarray}
To prove that this term vanishes uniformly in $t'$, we have to
show
 \beq \lim_{\la \downarrow 0} \sup_{0 \leq t' \leq t}
\int_0^{t'} \d s_1\int_{\la^{-2}(t'-s_1)}^{+\infty} \d u \, h(u)
=0.
 \eeq
This follows since for each $s_1 <t'$, the integral over $u$
vanishes as $\la \downarrow 0$ and  the whole expression is
bounded by $t' \|h \|_1$.

Since we have established that no pairing crosses the $t'$ point,
the problem factorizes and \eqref{eq: factorization} is true.
\end{proof}

\subsection{Convergence of annihilation operators}
\label{sec: preliminary lemmas}

\begin{lemma}\label{lem: conv of creation}
 For $\psi \in \caD$ and $j \in \caJ$,
\begin{eqnarray}
&&\lim_{\la \downarrow 0}  \la^{-1} a( J_\lambda^*
\e^{ \iu  \la^{-2} t (H_\res -\om(j)) } \phi_j )  \psi \\
&&   =        a \left(  \e^{ \iu t Z_\res   } |1 \rangle  \otimes
\phi_j(\om(j)) \right) \psi
\end{eqnarray}
uniformly in $t \in \bbR$.
\end{lemma}
\begin{proof}
We use
\begin{eqnarray} \label{eq: conv of anni} &&
\la^{-1}  J_\lambda^*
\e^{ \iu  \la^{-2} t (H_\res -\om(j)) } \phi_j (x)\\
&=&
\mathop{\oplus}\limits_{\omega\in\caF}
\la^{-1}  J_{\lambda,\omega}^*
\e^{ \iu  \la^{-2} t (H_\res -\om(j)) } \phi_j (x).\nonumber
\end{eqnarray}
Now
\begin{eqnarray*}&&
\la^{-1}  J_{\lambda,\omega}^*
\e^{ \iu  \la^{-2} t (H_\res -\om(j)) } \phi_j (x)\\
&=&\e^{\iu t(x+\lambda^{-2}(\omega-\omega(j)))}\left(1_{    I_\omega}(H_\res)\phi_j\right) (\omega+\lambda^2 x)\\
\\&\mathop{\longrightarrow}\limits_{\lambda\searrow0}
& \left\{\begin{array}{ll}\phi_j(\omega(j)),&\omega=\omega(j);\\
0,&\omega\neq\omega(j).\end{array}\right.
\end{eqnarray*}
\end{proof}

\subsection{Resummation formula}

The following lemma gives a convenient expression for the full
dynamics in terms of the reduced dynamics.

\begin{lemma}
\begin{eqnarray}
&&T_\lambda(t,t_0)\label{resummed_formula}\\
&=&\sum_{m=0}^\infty\int_{\triangle_m[t_0,t]}\d t_1\dots\d
t_m\sum_{\epsilon_1,\dots,\epsilon_p\in\{+,-\}}\sum_{j_1,\dots,j_m\in
  \caJ}(-\iu)^m \lambda^m\nonumber\\
&&\times I_\caK^* T_\lambda(t,t_m)D_{j_m}^{\epsilon_m}(t_m)T_\lambda
(t_m,t_{m-1})\cdots
D_{j_1}^{\epsilon_1}(t_1)T_\lambda(t_1,t_0)I_\caK\otimes
1_{\Gamma_\s(\caH_\res)}\nonumber\\
&&\times\prod_{i=1,\dots,m\ :\ \epsilon_i=+}a^*(\e^{\i t_i H_\res}\phi_{j_i})
\prod_{i'=1,\dots,m\ :\ \epsilon_{i'}=-}a(\e^{\i t_{i'}
  H_\res}\phi_{j_{i'}}).\nonumber
\end{eqnarray}
\end{lemma}

\begin{proof}
 $\widetilde{\Pair}(n)$ will denote the set of all pairings {\em
   inside} the set $\{1,\dots,n\}$. That means,
 $\sigma\in\widetilde{\Pair}(n)$ iff there is $p=0,1,\dots,[n/2]$ such that
$\sigma$ is an injection of $\{1,\dots,2p\}$ into $\{1,\dots,n\}$ satisfying
 $\sigma(2i-1)<\sigma(2i+1)$, $i=1,\dots,p-1$ and $\sigma(2i-1)<\sigma(2i)$,
 $i=1,\dots,p$.
For  $\sigma\in\widetilde{\Pair}(n)$, let $\Ran\sigma$ denote the image of
$\sigma$. We say that a sequence $\epsilon_1,\dots\epsilon_n$ of $\{+,-\}$ is
{\em compatible} with  $\sigma\in\widetilde{\Pair}(n)$  iff
\[\epsilon_{\sigma(2i-1)}=-,\ \ \epsilon_{\sigma(2i)}=+,\ \ i=1,\dots,p.\]
Applying the Dyson expansion and then the Wick theorem we obtain that the
left hand
side of (\ref{resummed_formula}) equals
\begin{eqnarray}&&
\sum_{n=0}^\infty\int_{\triangle_n[t_0,t]}\d t_1\cdots\d
t_n\sum_{j_1,\dots,j_n\in \caJ}
\sum_{\epsilon_1,\dots,\epsilon_n\in\{+,-\}}\nonumber
\\&&\times\prod_{r=1}^n
(-\iu)\lambda D_{j_r}^{\epsilon_r}(t_r)
\nonumber
\\&&\times
a^{\epsilon_n}(e^{\iu t_n H_\res}\phi_{j_n})\cdots
a^{\epsilon_1}(e^{\iu t_1 H_\res}\phi_{j_1})\nonumber\\&=&
\sum_{n=0}^\infty\int_{\triangle_n}\d t_1\cdots\d t_n\sum_{j_1,\dots,j_n\in \caJ}
\sum_{\sigma\in\widetilde{\Pair}(n)}
\sum_{\begin{array}{c}\scriptstyle
\epsilon_1,\dots,\epsilon_n\in\{+,-\}\\ \hbox{\tiny compatible with }\
\scriptscriptstyle \sigma\end{array} }\nonumber\\&&\times
\prod_{r=1}^n
(-\iu)\lambda D_{j_r}^{\epsilon_r}(t_r)
\nonumber\\&&\times
\prod_{i\in\{1,\dots,n\}\backslash\Ran\sigma\ :\ \epsilon_i=+}
a^*(e^{\iu t_iH_\res}\phi_{j_i})\prod_{i'\in\{1,\dots,n\}\backslash\Ran\sigma\ :\ \epsilon_{i'}=-}
a(e^{\iu t_{i'}H_\res}\phi_{j_{i'}})\nonumber\\&&\times
\prod_{q=1}^p\langle\phi_{j_{\sigma(2q)}}|\e^{\iu(t_{\sigma(2q)}-t_{\sigma(2q-1)})H_\res}
\phi_{j_{\sigma(2q-1)}}\rangle.\label{resummed_formula1}
\end{eqnarray}

Applying the Wick theorem and the Dyson expansion backwards we see that the
right hand side of (\ref{resummed_formula1}) equals the right hand side of
(\ref{resummed_formula}).
\end{proof}

\subsection{Proof of Theorem \ref{thm: stochastic limit}}

Set
\[D_{j,\omega}:=\sum_{e-e'=\omega}1_{\caK_e}D_j1_{\caK_{e'}}
.\]
Recall the operators $\nu_\omega$ and $\nu$ defined in
Subsection \ref{sec: asred}. They can be expressed in terms of
$D_{j,\omega}$ by
\begin{eqnarray*}
\nu_\omega&=& \wdr{\sqrt{2\pi}}\sum_{j\in \caJ\ :\
  \omega(j)=\omega}D_{j,\omega}\otimes|\phi_j(\omega)\rangle,\\
\nu&=& \wdr{\sqrt{2\pi}} \sum_{j\in \caJ}D_{j,\omega(j)}\otimes|\phi_j(\omega(j))\rangle.\\
\end{eqnarray*}

We use first the resummation formula (\ref{resummed_formula}), and then
we replace $D_j^\epsilon(t)$ with
\[\sum_{\omega\in\caF}D_{j,\omega}^\epsilon\e^{-\iu\epsilon\omega t}.\]
 We compute in terms of a quadratic form on $\caD$:
\begin{eqnarray}\label{expres}
&&\Gamma(J_\lambda^*)T_\lambda(\lambda^{-2}t,\lambda^{-2}t_0)\Gamma(J_\lambda)\\
 &=&\sum_{m=0}^\infty\ \ \
\int\limits_{\triangle_m[t_0,t]}\d t_1\dots\d
t_m\sum_{j_1,\dots,j_m}\sum_{\epsilon_1,\dots,\epsilon_m\in\{+,-\}}
(\iu\lambda)^{-m}
 \nonumber\\
&&\times I_\caK^* T_\lambda(\lambda^{-2}t,\lambda^{-2}t_m)D_{j_m}^{\epsilon_m}(\lambda^{-2}t_m)
T_\lambda(\lambda^{-2}t_m,\lambda^{-2}t_{m-1})\cdots\nonumber\\&&\times\cdots
D_{j_1}^{\epsilon_1}(\lambda^{-2}t_1)T_\lambda(\lambda^{-2}t_1,\lambda^{-2}t_0)
I_\caK\otimes 1_{\Gamma_\s(\caH_\res)}
\nonumber\\
&&\times\prod_{i=1,\dots,m\ :\ \epsilon_i=+}a^*(J_\lambda^*
\e^{\iu\lambda^{-2} t_i H_\res}\phi_{j_i})\nonumber\\&&\times
\Gamma(J_\lambda^*J_\lambda)
\prod_{i'=1,\dots,m\ : \ \epsilon_i=-}a(J_\lambda^*\e^{\iu\lambda^{-2} t_{i'}
  H_\res}\phi_{j_{i'}})\nonumber
\\
 &=&\sum_{m=0}^\infty\ \ \
\int\limits_{\triangle_m[t_0,t]}\d t_1\dots\d
t_m\sum_{j_1,\dots,j_m}\sum_{\epsilon_1,\dots,\epsilon_m\in\{+,-\}}
\ \ \ \sum_{\omega_1,\dots,\omega_m\in\caF}(\iu\lambda)^{-m}\nonumber
\\&&\times\prod_{p=1}^m
\e^{\iu(\omega_p-\omega(j_p))\lambda^{-2}t}
\nonumber\\
&&\times I_\caK^* T_\lambda(\lambda^{-2}t,\lambda^{-2}t_m)D_{j_m,\omega_m}^{\epsilon_m}
T_\lambda(\lambda^{-2}t_m,\lambda^{-2}t_{m-1})\cdots \nonumber\\&&\times\cdots
D_{j_1,\omega_1}^{\epsilon_1}T_\lambda(\lambda^{-2}t_1,\lambda^{-2}t_0)I_\caK\otimes
1_{\Gamma_\s(\caH_\res)}\nonumber\\
&&\times\prod_{i=1,\dots,m\ :\ \epsilon_i=+}a^*(J_\lambda^*
\e^{\i \lambda^{-2}t_i( H_\res-\omega(j_i))}\phi_{j_i})
\nonumber\\&&\times
\Gamma(J_\lambda^*J_\lambda)
\prod_{i'=1,\dots,m\ :\ \epsilon_i=-}
a(J_\lambda^*\e^{\iu\lambda^{-2} t_{i'}
 ( H_\res-\omega(j_{i'}))}\phi_{j_{i'}})\nonumber.
\end{eqnarray}
Now by Theorem \ref{thm: duemcke}, we have a uniform limit
\begin{eqnarray}
&&\lim_{\lambda\searrow0}
 I_\caK^* T_\lambda(\lambda^{-2}t,\lambda^{-2}t_m)D_{j_m,\omega_m}^{\epsilon_m}
T_\lambda(\lambda^{-2}t_m,\lambda^{-2}t_{m-1})\cdots\nonumber\\&&\times\cdots
D_{j_1,\omega_1}^{\epsilon_1}T_\lambda(\lambda^{-2}t_1,\lambda^{-2}t_0)I_\caK
\nonumber\\
&=&
\e^{-\iu(t-t_m)\Upsilon}D_{j_m,\omega_m}^{\epsilon_m}
\e^{-\iu(t_m-t_{m-1})\Upsilon}\cdots\nonumber\\&&\times\cdots
D_{j_1,\omega_1}^{\epsilon_1}\e^{\iu(t_1-t_0)\Upsilon}
\end{eqnarray}
By Lemma \ref{lem: conv of creation}, for $\psi,\psi'\in\caD$ we
have the uniform limits
\begin{eqnarray*}
&&\lim_{\lambda\searrow0}
\prod_{i'=1,\dots,m\ :\ \epsilon_{i'}=-}\lambda^{-1}
a(J_\lambda^*\e^{\iu\lambda^{-2} t_{i'}
 ( H_\res-\omega(j_{i'}))}\phi_{j_{i'}})\psi'\\
&=&
\prod_{i'=1,\dots,m\ :\ \epsilon_{i'}=-}
a\left(\e^{\iu t_{i'} Z_\res}
  |1\rangle\otimes\phi_{j_{i'}}(\omega(j_{i'}))\right)\psi' ,\\
&&\lim_{\lambda\searrow0}
\prod_{i=1,\dots,m\ :\ \epsilon_i=+}\lambda^{-1}
a(J_\lambda^*\e^{\iu\lambda^{-2} t_{i}
 ( H_\res-\omega(j_{i}))}\phi_{j_{i}})\psi\\
&=&
\prod_{i=1,\dots,m\ :\ \epsilon_i=+}
a\left(\e^{\iu t_{i} Z_\res}
  |1\rangle\otimes\phi_{j_{i}}(\omega(j_{i}))\right)\psi.
\end{eqnarray*}
Clearly,
$\s-\lim_{\lambda\searrow0}\Gamma(J_\lambda^*J_\lambda)=1$. Thus,
(\ref{expres}), as a quadratic form on $\caD$,
 up to an error of the order
$o(\lambda^0)$ equals
\begin{eqnarray}
&&\sum_{m=0}^\infty\ \
\int\limits_{\triangle_m[t_0,t]}\d t_1\dots\d
t_m\sum_{\epsilon_1,\dots,\epsilon_m\in\{+,-\}}\ \ \
\sum_{\omega_1,\dots,\omega_m\in\caF} (-\iu)^m \nonumber\\
&&\times\prod_{p=1}^m
\e^{\iu(\omega_p-\omega(j_p))\lambda^{-2}t}
 \nonumber\\
&&\times \e^{-\iu(t-t_m)\Upsilon}D_{j_m,\omega_m}^{\epsilon_m}
\e^{-\iu(t_m-t_{m-1})\Upsilon}\cdots\nonumber\\&&\times\cdots
D_{j_1,\omega_1}^{\epsilon_1}\e^{-\iu(t_1-t_0)\Upsilon}
\otimes 1_{\Gamma_\s(\caH_\res)}
\nonumber\\&&\times
\prod_{i=1,\dots,m\ :\ \epsilon_{i}=+}
a\left(\e^{\iu t_{i} Z_\res}
  |1\rangle\otimes\phi_{j_{i}}(\omega(j_{i}))\right)\nonumber\\&&\times
\prod_{i'=1,\dots,m\ :\ \epsilon_{i'}=-}
a\left(\e^{\iu t_{i'} Z_\res}
  |1\rangle\otimes\phi_{j_{i'}}(\omega(j_{i'}))\right). \nonumber
\end{eqnarray}
By the Riemann-Lebesgue Lemma, in the limit $\lambda\searrow0$,
all the terms with $\omega(j_p)\neq\omega_p$ for some $p$
disappear, and we obtain
\begin{eqnarray}
&&\sum_{m=0}^\infty\ \
\int\limits_{\triangle_m[t_0,t]}\d t_1\dots\d
t_m\sum_{\epsilon_1,\dots,\epsilon_m\in\{+,-\}} (-\iu)^m \nonumber\\
&&\times \e^{-\iu(t-t_m)\Upsilon}D_{j_m,\omega_m}^{\epsilon_m}
\e^{-\iu(t_m-t_{m-1})\Upsilon}\cdots\nonumber\\&&\times\cdots
D_{j_1,\omega_1}^{\epsilon_1}\e^{-\iu(t_1-t_0)\Upsilon}
\otimes 1_{\Gamma_\s(\caH_\res)}
\nonumber\\&&\times
\prod_{i=1,\dots,m\ :\ \epsilon_{i}=+}
a\left(\e^{\iu t_{i} Z_\res}
  |1\rangle\otimes\phi_{j_{i}}(\omega(j_{i}))\right)\nonumber\\&&\times
\prod_{i'=1,\dots,m\ :\ \epsilon_{i'}=-}
a\left(\e^{\iu t_{i'} Z_\res}
   |1\rangle\otimes\phi_{j_{i'}}(\omega(j_{i'}))\right) \nonumber
\\&=&\e^{\iu t\d\Gamma(Z_\res)}U_{t-t_0}
\e^{-\iu t_0\d\Gamma(Z_\res)}.\nonumber
\end{eqnarray}
Thus we obtained, for $\psi,\psi'\in\caD$,
\begin{eqnarray}\label{limmi}
&& \lim_{\la \downarrow 0}
\langle\psi|\Ga  (J^*_{\la} )
T_\lambda(\lambda^{-2}t,\lambda^{-2}t_0)
\Gamma(J_\lambda)\psi'\rangle\\& &=
\langle\psi|
\e^{\iu t \d \Ga (Z_\res)} U_{t-t_0}\e^{-\iu t_0 \d \Ga (Z_\res)} \psi'  \rangle.
\end{eqnarray}
By density, (\ref{limmi}) can be extended to the weak limit on the whole
space. But the weak convergence of contractions to a unitary operator implies
the strong* convergence. This  yields \eqref{thm: stochastic limit 2}.

Note that
\[\Gamma(J_\la^*)\e^{-\i t\la^{-2}tH_0}(1-\Gamma\left(J_\la
J_\la^*)\right)=0.\]
Therefore,
\begin{eqnarray}
& \e^{\iu \la^{-2} t Z_\ren}  \Gamma (J_{\la}^* ) \e^{-\iu
\la^{-2} t H_{\la}}\Gamma (J_{\la} ) & \\
= & \e^{\iu \la^{-2} t Z_\ren} \Gamma (J_{\la}^* ) \e^{-\iu
\la^{-2} t H_{0}} \Gamma
(J_{\la})   &   \nonumber \\
 &  \times \Gamma
(J_{\la}^{*}) \e^{\iu \la^{-2} t H_{0}}
  \e^{-\iu \la^{-2} t H_{\la}} \Gamma (J_\la ).       &  \nonumber
\end{eqnarray}
The equality \eqref{thm: stochastic limit 1} now follows from
\eqref{thm: stochastic limit 3} and \eqref{thm: stochastic limit
2} since the strong limit of a product of uniformly bounded
operators is the product of limits.

\subsection{Proof of Theorem \ref{thm: algebra}}

Remark that \beq  \label{eq: almost unitarity} (1-\Gamma (J_\la
J_{\la}^{*})) \e^{\iu \la^{-2} t H_{0}}
  \e^{-\iu \la^{-2} t H_{\la}} \Gamma (J_\la ) \mathop{\longrightarrow}\limits_{\la \searrow 0}^{\mathrm{strongly}} 0 . \eeq

This follows since for $\psi \in \caZ$
  \begin{eqnarray*}
&&
\| (1-\Gamma (J_\la
J_{\la}^{*})) \e^{\iu \la^{-2} t H_{0}}
  \e^{-\iu \la^{-2} t H_{\la}} \Gamma (J_\la ) \psi\|^2\\
&=&
\|  \e^{\iu \la^{-2} t H_{0}}
  \e^{-\iu \la^{-2} t H_{\la}} \Gamma (J_\la ) \psi\|^2
-\| \Gamma (J_\la
J_{\la}^{*}) \e^{\iu \la^{-2} t H_{0}}
  \e^{-\iu \la^{-2} t H_{\la}} \Gamma (J_\la ) \psi\|^2\\
&=&
\| \Gamma (J_\la ) \psi\|^2
-\| \Gamma (J_{\la}^{*}) \e^{\iu \la^{-2} t H_{0}}
  \e^{-\iu \la^{-2} t H_{\la}} \Gamma (J_\la ) \psi\|^2
 \mathop{\longrightarrow}\limits_{\la \searrow 0} 0  ,
  \end{eqnarray*}
where we used  $\slim\limits_{\la\searrow0}J_\la^*J_\la=1$, Theorem \ref{thm: stochastic limit} and the fact
that $\Ga  (J_\la)$ is a partial isometry.

Theorem \ref{thm: algebra} is now proven by using \eqref{thm:
stochastic limit 2}, \eqref{connection Theta J}, \eqref{eq: almost
unitarity}  and the fact that for multiplication operators ${G}$
as in the text preceding Theorem \ref{thm: algebra} we have
\[  [\e^{\iu t \d \Ga (H_\res ) }, {\Ga(G)} ] =0  \qquad  [ {\Ga(J_\la
    J^*_\la)} , {\Ga(G)} ] =0.     \]

\bibliographystyle{plain}
\bibliography{paulifierz}

\end{document}